\documentclass[12pt]{article}
\pdfoutput=1

\usepackage{color}
\usepackage{graphicx}
\usepackage{amsmath}
\usepackage{slashed}
\usepackage{amssymb}
\usepackage{epstopdf}

\input xy
\xyoption{all}
\xyoption{web}

\usepackage[
      colorlinks=true,
      linkcolor=blue,
      urlcolor=blue,
      filecolor=black,
      citecolor=red,
      pdfstartview=FitV,
      pdftitle={},
        pdfauthor={Yi Li},
        pdfsubject={},
        pdfkeywords={},
        pdfpagemode={},
        bookmarksopen=true
      ]{hyperref}

\usepackage{color}
\usepackage{graphicx}

\usepackage{sectsty}
\sectionfont{\large}

\renewcommand{\thefootnote}{\fnsymbol{footnote}}
\renewcommand{\thanks}[1]{\footnote{#1}}
\newcommand{\starttext}{
\setcounter{footnote}{0}
\renewcommand{\thefootnote}{\arabic{footnote}}}

\newcommand{\bea}{\begin{eqnarray}}
\newcommand{\eea}{\end{eqnarray}}
\newcommand{\be}{\begin{equation}}
\newcommand{\ee}{\end{equation}}
\newcommand{\<}{\langle}
\renewcommand{\>}{\rangle}

\DeclareMathOperator{\csch}{csch}

\usepackage{ dsfont } 

\long\def\symbolfootnote[#1]#2{\begingroup%
\def\thefootnote{\fnsymbol{footnote}}\footnote[#1]{#2}\endgroup}

\begin{document}
\setlength{\baselineskip}{18pt}

\starttext
\setcounter{footnote}{0}

%
\bigskip

\begin{center}

{\Large \bf  Cutoff $\rm AdS_3$ versus $\rm T\bar{T}$ $\rm CFT_2$ in the large central charge sector: correlators of energy-momentum tensor}

\vskip 0.4in

{\large Yi Li and Yang Zhou}

\vskip .2in

{\it Department of Physics and Center for Field Theory and Particle Physics}\\
{\it Fudan University, Shanghai 200433, China}\\[0.5cm]
\href{mailto:liyi@fudan.edu.cn}{\texttt{liyi@fudan.edu.cn}}\texttt{, }\href{mailto:yang_zhou@fudan.edu.cn}{\texttt{yang\_zhou@fudan.edu.cn}}

\bigskip

\bigskip

\end{center}

\begin{abstract}

\setlength{\baselineskip}{18pt}

In this article we probe the proposed holographic duality between $T\bar{T}$ deformed two dimensional conformal field theory and the gravity theory of $\rm AdS_3$ with a Dirichlet cutoff by computing correlators of energy-momentum tensor. We focus on the large central charge sector of the $T\bar{T}$ CFT in a Euclidean plane and a sphere, and compute the correlators of energy-momentum tensor using an operator identity promoted from the classical trace relation. The result agrees with a computation of classical pure gravity in Euclidean $\rm AdS_3$ with the corresponding cutoff surface, given a holographic dictionary which identifies gravity parameters with $T\bar{T}$ CFT parameters.
\end{abstract}

\setcounter{equation}{0}
\setcounter{footnote}{0}

%
%
%
%
%
\newpage


\section{Introduction}
\setcounter{equation}{0}
\label{sec1}
The $T\bar{T}$ deformation of two dimensional quantum field theory has received intensive study in the past few years. As an irrelevant deformation, it leads to well-defined, albeit non-local, UV completion. In fact, it is a solvable deformation in many senses. It preserves integrability structures \cite{Smirnov:2016lqw}\cite{Conti:2018jho}, deforms the scattering matrix by multiplying CDD factors \cite{Cavaglia:2016oda}\cite{Dubovsky:2017cnj}, has solvable deformation of finite size spectrum \cite{Cavaglia:2016oda}\cite{Zamolodchikov:2004ce} and preserves modular invariance of conformal field theory torus partition function \cite{Datta:2018thy}\cite{Aharony:2018bad}. The non-locality and solvability of the $T\bar{T}$ deformation can be understood from a different perspective by reformulation to random geometry \cite{Cardy:2018sdv}, which also neatly derives the flow equation of the partition function. In addition, the $T\bar{T}$ deformation can be re-interpreted as coupling to Jackiw-Teitelboim gravity of the quantum field theory, which leads to the same flow equation of the partition function and CDD factors of the scattering matrix \cite{Dubovsky:2017cnj}\cite{Dubovsky:2018bmo}. Correlators of $T\bar{T}$ deformed QFT or CFT were studied in \cite{Cardy:2019qao}\cite{He:2019vzf}\cite{He:2020udl}. While much of the work on the $T\bar{T}$ deformation has been done in the flat Euclidean plane or its quotient spaces such as cylinder and torus,  generalization to maximally symmetric spaces was considered in \cite{Jiang:2019tcq}\cite{Brennan:2020dkw}. Further generalization to generic curved spaces was studied in \cite{Tolley:2019nmm}\cite{Mazenc:2019cfg}, which has remarkably reproduced lots of result of previous study.

For a holographic $\rm CFT_2$, it's natural to ask what the holographic dual of its $T\bar{T}$ deformation is. It was proposed by Mezei et al. \cite{McGough:2016lol} that for positive $T\bar{T}$ deformation parameter the holographic dual is a Dirichlet cutoff in the $\rm AdS_3$ gravity, based on computation of signal propagation speed, quasi-local energy of BTZ blackhole and other physics quantities. It was followed by study on holographic entanglement entropy  \cite{Donnelly:2018bef}\cite{Chen:2018eqk}\cite{Banerjee:2019ewu}\cite{Jeong:2019ylz}\cite{Grieninger:2019zts}\cite{Lewkowycz:2019xse}\cite{Geng:2019ruz}\cite{Donnelly:2019pie},  generalization to higher or lower dimensions \cite{Bonelli:2018kik}\cite{Taylor:2018xcy}\cite{Hartman:2018tkw}\cite{Caputa:2019pam}\cite{Gross:2019ach}\cite{Gross:2019uxi}\cite{Iliesiu:2020zld}, and an interesting perspective from path integral optimization \cite{Jafari:2019qns}. In addition, the proposal was examined by holographic computation of correlators of energy-momentum tensor in \cite{Kraus:2018xrn}. It was found that the large central charge perturbative correlators in $T\bar{T}$ $\rm CFT_2$ agree with correlators of classical pure gravity in cutoff $\rm AdS_3$ given a holographic dictionary that identifies gravity parameters with $T\bar{T}$ CFT parameters. But additional non-local double trace deformation must be supplemented to the $T\bar{T}$ deformation to reproduce correlators of scalar operators dual to matter fields added to gravity, in line with the general discussion of bulk cutoff in \cite{Heemskerk:2010hk}\cite{Faulkner:2010jy}. The possible limitation of the Dirichlet cutoff picture was echoed in \cite{Guica:2019nzm}, which showed that in the large central charge limit the holographic dual of $T\bar{T}$ $\rm CFT_2$ in the Euclidean plane is in general $\rm AdS_3$ gravity with mixed boundary condition, and only for positive deformation parameter and for pure gravity the mixed boundary condition can be reinterpreted as Dirichlet boundary condition at a finite cutoff, taking the original form proposed by Mezei et al..

This article is to a large extent a follow-up of \cite{Kraus:2018xrn}, and \cite{Aharony:2018vux} which computed the correlators of energy-momentum tensor of $T\bar{T}$ CFT in a Euclidean plane beyond leading order in the large central charge limit. We start in Section 2 by a brief review of $T\bar{T}$ deformation which highlights a trace relation formula. In Section 3 we promote the trace relation to an operator identity and compute in the large central charge limit the correlators of energy-momentum tensor for $T\bar{T}$ CFT in a Euclidean plane, a sphere and a hyperbolic space. In Section 4 we compute correlators of energy-momentum tensor in classical pure gravity in Euclidean $\rm AdS_3$ cut off by a Euclidean plane and a sphere. The gravity correlators are found to agree with $T\bar{T}$ CFT correlators given a dictionary between $T\bar{T}$ CFT parameters and gravity parameters. In Section 5 we summarize our result and discuss  related questions and possible directions of further research.

\section{$T\bar{T}$ deformation and trace relation}
\setcounter{equation}{0}
\label{sec2}
The $T\bar{T}$ deformation with the continuous deformation parameter $\mu$ is defined by a flow of action in the direction of $T\bar{T}$ operator
\begin{align} \label{TTbarDeformationDefinition}
 \frac{d S}{d\mu} = \int dV T\bar{T}
\end{align}
The $T\bar{T}$ operator is a covariant quadratic combination of energy-momentum tensor
\footnote{Here we follow the normalization of $T\bar{T}$ operator in \cite{McGough:2016lol} and \cite{Donnelly:2018bef}.}
\begin{align} 
 T\bar{T} = \frac{1}{8}(T^{ij}T_{ij}-{T^i_i}^2)
\end{align}
where the energy-momentum tensor is defined in the convention
\begin{align} \label{EMtensorConvention}
 \delta S = \frac{1}{2} \int dV T^{ij} \delta g_{ij}
\end{align}
It was shown in \cite{Zamolodchikov:2004ce} that the composite $T\bar{T}$ operator has an unambiguous and UV finite definition modulo derivative of local operators by limit of point splitting
\begin{align} \label{TTbarOperatorFromPointSplitting}
 T\bar{T}(x) = \lim_{y\rightarrow x} \frac{1}{8}(T^{ij}(x)T_{ij}(y)-T^i_i(x)T^j_j(y))
\end{align}
for quantum field theory in the Euclidean plane with a conserved and symmetric energy-momentum tensor.  This point splitting definition can be generalized to maximally symmetric spaces by carrying over Zamolodchikov's argument, but it was found that the factorization property of the expectation value
\begin{align} \label{TTbarFactorization}
 \<T\bar{T}\> = \frac{1}{8}(\<T^{ij}\>\<T_{ij}\>-\<{T^i_i}\>^2)
\end{align}
is lost in general \cite{Zamolodchikov:2004ce}\cite{Jiang:2019tcq}.

We refer interested readers to Jiang's note  \cite{Jiang:2019hxb} and other references for many interesting properties of $T\bar{T}$ CFT. Here we focus on the trace relation crucial for computation in the following sections
\begin{align} \label{TTbarTraceRelation}
 T^i_i = -2\mu T\bar{T}
\end{align}
When regarded as a classical field equation it was discovered in free scalar theory \cite{Cavaglia:2016oda}, and was later proved for $T\bar{T}$ $\rm CFT_2$ in generic curved spaces in \cite{Tolley:2019nmm}. Actually, we have a very basic argument for theories with Lagrangian density $\cal L$ as an algebraic function of the metric.
\footnote{Free scalar falls into this category.}
For these theories, the energy-momentum tensor takes the form
\begin{align}
 T_{ij} = g_{ij} {\cal L} - 2 \frac{\partial{\cal L}}{\partial g^{ij}}
\end{align}
and we have the $T\bar{T}$ flow equation for the Lagrangian density
\begin{align} \label{TTbarLagrangianFlowEqn}
 \partial_\mu {\cal L} &= T\bar{T} = \frac{1}{8}(T^{ij}T_{ij}-{T^i_i}^2) \nonumber\\
 &= \frac{1}{4}(-{\cal L}^2 + 2{\cal L}g^{ij}\frac{\partial {\cal L}}{\partial g^{ij}} + 4g^{ik}g^{jl}\frac{\partial {\cal L}}{\partial g^{ij}}\frac{\partial{\cal L}}{\partial g^{kl}} - 4g^{ij}g^{kl}\frac{\partial {\cal L}}{\partial g^{ij}}\frac{\partial{\cal L}}{\partial g^{kl}})
\end{align}
And the trace relation takes the form
\begin{align}
 \mu \partial_\mu {\cal L} + {\cal L} - g^{ij} \frac{\partial{\cal L}}{\partial g^{ij}}=0
\end{align}
Taking derivative of the left hand side of the equation above with respect to $\mu$ and using (\ref{TTbarLagrangianFlowEqn}) we get
\begin{align}
 & \partial_\mu(\mu \partial_\mu {\cal L} + {\cal L} - g^{ij} \frac{\partial{\cal L}}{\partial g^{ij}}) = -\frac{1}{2}{\cal L}(\mu \partial_\mu {\cal L} + {\cal L} - g^{ij} \frac{\partial{\cal L}}{\partial g^{ij}}) + \frac{1}{2}g^{ij}\frac{\partial {\cal L}}{\partial g^{ij}}(\mu \partial_\mu {\cal L} + {\cal L} - g^{mn} \frac{\partial{\cal L}}{\partial g^{mn}}) \nonumber\\
 &+ \frac{1}{2}{\cal L}g^{ij} \frac{\partial}{\partial g^{ij}}(\mu \partial_\mu {\cal L} + {\cal L} - g^{mn} \frac{\partial{\cal L}}{\partial g^{mn}}) + 2 g^{ik}g^{jl} \frac{\partial}{\partial g^{ij}}(\mu \partial_\mu {\cal L} + {\cal L} - g^{mn} \frac{\partial{\cal L}}{\partial g^{mn}})\frac{\partial{\cal L}}{\partial g^{kl}} \nonumber\\
 &- 2 g^{ij}g^{kl} \frac{\partial}{\partial g^{ij}}(\mu \partial_\mu {\cal L} + {\cal L} - g^{mn} \frac{\partial{\cal L}}{\partial g^{mn}})\frac{\partial{\cal L}}{\partial g^{kl}}
\end{align}
The trace relation holds at $\mu=0$ as a paraphrase that the energy-momentum tensor in CFT is traceless. By the first order differential equation above it must hold for all $\mu$. For quantum theory we expect quantum corrections to the trace relation, it depends on how $T\bar{T}$ deformation is defined for quantum field theory in curve spaces. 
\footnote{It takes the form of Wheeler-de Witt equation in the scheme of $T\bar{T}$ in curved spaces as quantum 3D gravity in \cite{Mazenc:2019cfg}.}
In our work we assume it holds as an operator identity within connected correlators, at least in the large central charge limit, and the $T\bar{T}$ operator is given by the point splitting definition since we work in maximally symmetric spaces.

\section{Correlators of energy-momentum tensor of $T\bar{T}$ deformed $\rm CFT_2$ in the large central charge limit}
\setcounter{equation}{0}
\label{sec3}
In this section we use the trace relation (\ref{TTbarTraceRelation}) to compute the correlators of energy-momentum tensor in the large central charge limit, a limit of large degrees of freedom similar to the large $N$ limit in gauge theory. More precisely it's a limit with a large central charge $c$ of the undeformed CFT, but finite $\mu c$ where $\mu$ is the $T\bar{T}$ deformation parameter. A detailed discussion of the large $c$ limit can be found in \cite{Aharony:2018vux}. Inspired by the work in \cite{Kraus:2018xrn} and \cite{Aharony:2018vux}, we first compute up to four point correlators of energy-momentum tensor for $T\bar{T}$ CFT in the two dimensional Euclidean plane $\mathbb{E}_2$. Then we consider $T\bar{T}$ CFT in the two dimensional sphere $\mathbb{S}_2$ and the two dimensional hyperbolic space $\mathbb{H}_2$ to compute up to three point correlators.

\subsection{Large $c$ correlators of $T\bar{T}$ CFT in $\mathbb{E}_2$}
In principle, our tools to compute correlators of energy-momentum tensor in this section are the trace relation, the conservation equation, dimensional analysis, Bose symmetry, CFT limit and other physical considerations. The conservation equation of energy-momentum tensor is
\begin{align} \label{Tconservation}
 \nabla^i T_{ij} = 0
\end{align}
It holds in a correlator except for contact terms. In the Euclidean plane the metric takes the form
\begin{align} \label{MetricE2}
 ds^2= dz d\bar{z}
\end{align}
in the complex coordinates $z,\bar{z}$ and the conservation equation is
\begin{align} \label{TconservationE2}
 &\partial_{\bar{z}} T_{zz} + \partial_z T_{z\bar{z}} = 0 \nonumber\\
 &\partial_{\bar{z}} T_{z\bar{z}} + \partial_z T_{\bar{z}\bar{z}} = 0
\end{align}
We have vanishing one point correlator
\begin{align} \label{1ptCorrelatorTTE2}
 \<T_{ij}\> = 0
\end{align}
and it's shown in \cite{Aharony:2018vux} that two point correlators remain the same as in the undeformed CFT in the large $c$ limit
\footnote{Here the superscript $(0)$ on $T$ indicates it's the energy-momentum tensor in the undeformed CFT, and the superscript $(0)$ on the expectation value means it's evaluated in the undeformed CFT, for example, by path integral with the undeformed CFT action. By this convention we should add superscript like $(\mu)$ for the energy-momentum tensor and the expectation value in the $T\bar{T}$ deformed CFT with deformation parameter $\mu$, but we choose to omit it for simplicity of the text.}
\footnote{For simplicity we omit correlators that can be simply inferred by symmetry, e.g. $\<T_{\bar{z}\bar{z}}(\vec{w})T_{\bar{z}\bar{z}}(\vec{v})\> = \frac{c}{8\pi^2}\frac{1}{(\bar{w}-\bar{v})^4}$.}
\footnote{A bit abuse of notation, we use the equality sign even if it's only equal in the large c limit, because we exclusively work in this limit.}
\begin{align} 
 &\<T_{zz}(w)T_{zz}(v)\> = \<T^{(0)}_{zz}(\vec{w})T^{(0)}_{zz}(\vec{v})\>^{(0)} = \frac{c}{8\pi^2}\frac{1}{(w-v)^4} \nonumber\\
 &\<T_{zz}(w)T_{z\bar{z}}(v)\> = \<T^{(0)}_{zz}(\vec{w})T^{(0)}_{z\bar{z}}(\vec{v})\>^{(0)} = 0
\end{align} 
It's sometimes convenient to use the normalization of energy-momentum tensor in CFT
\begin{align} 
 T=2\pi T_{zz},\quad \bar{T}=2\pi T_{\bar{z}\bar{z}},\quad \Theta=2\pi T_{z\bar{z}}
\end{align}
and the two point correlators now take the form
\begin{align} \label{2ptCorrelatorTTE2}
 &\<T(\vec{w})T(\vec{v})\> = \frac{c}{2}\frac{1}{(w-v)^4} \nonumber\\
 &\<T(\vec{w})\Theta(\vec{v})\> = 0
\end{align}
To compute the three point correlators, we start with $\<T(\vec{w})\Theta(\vec{v})\bar{T}(\vec{u})\>^c$ where the superscript $c$ means connected correlators. 
\footnote{In the Euclidean plane, two and three point correlators are equal to the connected counterparts because one point correlator vanishes.}
Using the trace relation \ref{TTbarTraceRelation} in the Euclidean plane
\begin{align} \label{TraceRelationE2}
 T_{z\bar{z}} = -\frac{\mu}{2}(T_{zz}T_{\bar{z}\bar{z}}-T_{z\bar{z}}^2)
\end{align}
or
\begin{align}
 \Theta(z) = -\frac{\mu}{4\pi}(T(z) \bar{T}(z)-\Theta(z)^2)
\end{align}
we get
\begin{align}
 \<T(\vec{w})\Theta(\vec{v})\bar{T}(\vec{u})\>^c = -\frac{\mu}{4\pi}\<T(\vec{w})(T(\vec{v})\bar{T}(\vec{v})-\Theta(\vec{v})^2)\bar{T}(\vec{u})\>^c
\end{align}
Working in the large c limit in which connected correlators of energy-momentum tensor scale as $c$, the correlator on the right hand side only contribute in the large $c$ limit by factorization into two correlators
\begin{align}
 \<T(\vec{w})\Theta(\vec{v})\bar{T}(\vec{u})\>^c == -\frac{\mu}{4\pi}\<T(\vec{w})T(\vec{v})\>^c\<\bar{T}(\vec{v})\bar{T}(\vec{u})\>^c = -\frac{\mu c^2}{16\pi} \frac{1}{(w-v)^4(\bar{v}-\bar{u})^4}
\end{align}
By the conservation equation $\partial_z \Theta + \partial_{\bar{z}} T=0$, we get $\<T(\vec{w})T(\vec{v})\bar{T}(\vec{u})\>^c = -\frac{\mu c^2}{12\pi} (\frac{1}{(w-v)^5(\bar{v}-\bar{u})^3}+(w \leftrightarrow v))$ modulo a holomorphic function in $\vec{v}$. By Bose symmetry it must be holomorphic in $\vec{w}$ as well, then it cannot depend on $\vec{u}$ at all by translational symmetry, and it's further fixed to be zero by cluster decomposition principle. Other correlators can also be computed in this way except for $\<T(\vec{w})T(\vec{v})T(\vec{u})\>^c$ and $\<\bar{T}(\vec{w})\bar{T}(\vec{v})\bar{T}(\vec{u})\>^c$, we only know $\<T(\vec{w})T(\vec{v})T(\vec{u})\>^c$ is holomorphic by the conservation equation and it has the CFT limit $c \frac{1}{(w-v)^2(v-u)^2(u-w)^2}$. However, it was proved in \cite{Aharony:2018vux} that $n$ point correlators are polynomial in $\mu$ of degree $n-2$, so we can rule out possible additional terms dependent on $\mu$ like $\mu^3 c^4\frac{1}{(w-v)^4(v-u)^4(u-w)^4}$. To summarize we list non-zero three point correlators
\begin{align} \label{3ptCorrelatorTTE2}
 &\<T(\vec{w})T(\vec{v})T(\vec{u})\>^c = c \frac{1}{(w-v)^2(v-u)^2(u-w)^2} \nonumber\\
 &\<T(\vec{w})\Theta(\vec{v})\bar{T}(\vec{u})\>^c = -\frac{\mu c^2}{16\pi} \frac{1}{(w-v)^4(\bar{v}-\bar{u})^4} \nonumber\\
 &\<T(\vec{w})T(\vec{v})\bar{T}(\vec{u})\>^c = -\frac{\mu c^2}{12\pi} (\frac{1}{(w-v)^5(\bar{v}-\bar{u})^3}+(w \leftrightarrow v))
\end{align}
Compared to previous work a clarification is needed. This result has been obtained in \cite{Kraus:2018xrn} as the leading order in $\mu$ result, by using the trace relation to the leading order in $\mu$. Later in \cite{Aharony:2018vux} it was derived for $T\bar{T}$ free scalars as large $c$ result, that is, $c$ times arbitrary function of $\mu c$. Here we derive it as large $c$ result without assuming the specific model of the undeformed CFT, but we have to assume the operator identity promoted from the trace relation. In a similar way, we computed two four point correlators
\begin{align} \label{4ptCorrelatorTTE2}
 &\<\bar{T}(\vec{\zeta})\Theta(\vec{z})T(\vec{w})T(\vec{v})\>^c = -\frac{\mu}{4}\<\bar{T}(\vec{\zeta})(T(\vec{z})\bar{T}(\vec{z})-\Theta(\vec{z})^2)T(\vec{w})T(\vec{v})\>^c \nonumber\\
 &= -\frac{\mu}{4}(\<\bar{T}(\vec{\zeta})\bar{T}(z)\>^c \<T(\vec{z})T(\vec{w})T(\vec{v})\>^c + \<\bar{T}(\vec{\zeta})\bar{T}(\vec{z})T(\vec{w})\>^c\<T(\vec{z})T(\vec{v})\>^c + \<\bar{T}(\vec{\zeta})\bar{T}(\vec{z})T(\vec{v})\>^c\<T(\vec{z})T(\vec{w})\>^c) \nonumber\\
 &= -\frac{\mu c^2}{8 \pi} \frac{1}{(\bar{\zeta}-\bar{z})^4(z-w)^2(w-v)^2(v-z)^2} \nonumber\\
 & + \frac{\mu^2 c^3}{96 \pi^2}(\frac{1}{(z-v)^4(\bar{\zeta}-\bar{z})^5(z-w)^3}+\frac{1}{(z-v)^4(\bar{z}-\bar{\zeta})^5(\zeta-w)^3} + (w \leftrightarrow v)) \nonumber\\
 &\<\Theta(\vec{\zeta})\Theta(\vec{z})T(\vec{w})T(\vec{v})\>^c = \frac{\mu^2}{16} \<(T(\vec{\zeta})\bar{T}(\vec{\zeta})-\Theta(\vec{\zeta})^2)(T(\vec{z})\bar{T}(\vec{z})-\Theta(\vec{z})^2)T(\vec{w})T(\vec{v})\>^c \nonumber\\
 &= \frac{\mu^2}{16 \pi^2} (\<\bar{T}(\zeta)\bar{T}(z)\>^c\<T(\zeta)T(w)\>^c\<T(z)T(v)\>^c + (w \leftrightarrow v)) \nonumber\\
 &= \frac{\mu^2 c^3}{128 \pi^2}(\frac{1}{(\bar{\zeta}-\bar{z})^4(\zeta-w)^4(z-v)^4} + (w \leftrightarrow v))
\end{align}
One can continue in this procedure to obtain all higher point correlators.

\subsection{Large $c$ correlators of $T\bar{T}$ CFT in $\mathbb{S}^2$ and $\mathbb{H}^2$}
Now we study correlators of $T\bar{T}$ CFT in a two dimensional sphere of radius $r$ or a hyperbolic space of radius $r$. In a maximally symmetric space, one point correlator of energy-momentum tensor is proportional to the metric
\begin{align} 
 \<T_{ij}\>=\alpha g_{ij}
\end{align}
The coefficient can be determined by the trace relation in vacuum expectation value supplemented by a trace anomaly term \cite{McGough:2016lol}\cite{Donnelly:2018bef}, and by using large $c$ factorization, we get
\begin{align} \label{TTLargec}
 \<T^i_i\> = -\frac{\mu}{4}\<T^{ij}T_{ij}-(T^k_k)^2\> - \frac{c}{24\pi}R = -\frac{\mu}{4}(\<T^{ij}\>\<T_{ij}\>-\<T^k_k\>^2)- \frac{c}{24\pi}R
\end{align}
For sphere with radius $r$ the scalar curvature is $R=\frac{2}{r^2}$, we find
\begin{align} \label{1ptCorrelatorTTS2}
 \<T_{ij}\> = \frac{2}{\mu}(1-\sqrt{1+\frac{\mu c}{24\pi r^2}})g_{ij}
\end{align}
For hyperbolic space with radius $r$ the scalar curvature is $R=-\frac{2}{r^2}$, we find
\begin{align} \label{1ptCorrelatorTTH2}
 \<T_{ij}\> = \frac{2}{\mu}(1-\sqrt{1-\frac{\mu c}{24\pi r^2}})g_{ij}
\end{align}
We note a square root singularity occurs at $\mu = \frac{24\pi r^2}{c}$.

Higher point correlators are a bit more complicated in a curved space. They are multi-point tensors based on the (co)tangent spaces at those points. Because the sphere and the hyperbolic space are maximally symmetric, two point correlators must be maximally symmetric bi-tensors, that is, bi-tensors covariant with the isometry group. Maximally symmetric bi-tensor has been studied in \cite{Allen:1985wd} exactly in the context of tensorial two point correlators, and it has already been used in \cite{Osborn:1999az} to study correlators of energy-momentum tensor in maximally symmetric spaces. Recently it was reviewed in \cite{Jiang:2019tcq} to study expectation value of $T\bar{T}$ operator in maximally symmetric spaces in general dimensions. Following their analysis and assuming the energy-momentum tensor is traceless in connected correlators in the undeformed CFT, we get two point correlators of undeformed CFT in $\mathbb{S}^2$ and $\mathbb{H}^2$. Details of computation are left to the Appendix \ref{appa}. Two point correlators of energy-momentum tensor of CFT in $\mathbb{S}^2$ take the form
\begin{align} \label{2ptCorrelatorCFTS2}
 \<T^{(0)}(\vec{\rm w})T^{(0)}(\vec{\rm v})\>^{(0)c} = \frac{c}{2} \frac{1}{\rm (w-v)^4}
\end{align}
in the complex stereographic projection coordinates of the sphere
\footnote{We are using similar normalization as in $\mathbb{E}^2$, that is, $T=2\pi T_{\rm zz},\quad \bar{T}=2\pi T_{\rm \bar{z}\bar{z}},\quad \Theta=2\pi T_{\rm z\bar{z}}$.}
,in which the metric is
\begin{align} \label{MetricS2zzb}
 ds^2 = \frac{r^2 d{\rm z} d\bar{\rm z}}{\rm (1+\frac{z\bar{z}}{4})^2}
\end{align}
It's related to the spherical coordinates by ${\rm z}=2 \cot\frac{\theta}{2}\mathrm{e}^{i \phi} \quad \bar{\rm z}=2 \cot\frac{\theta}{2}\mathrm{e}^{-i \phi}$.
\footnote{Similar to the spherical coordinates, the stereographic projection coordinate patch misses one point of the sphere. That's remedied by imposing appropriate regularity condition of physics quantities as $|\rm z|\rightarrow \infty$.}
And two point correlators of energy-momentum tensor of CFT in $\mathbb{H}^2$ take the form
\begin{align} \label{2ptCorrelatorCFTH2}
 \<T^{(0)}(\vec{\rm w})T^{(0)}(\vec{\rm v})\>^{(0)c} = \frac{c}{2} \frac{1}{\rm (w-v)^4}
\end{align}
in the complex Poincare disk coordinates of the hyperbolic space, in which the metric is
\begin{align} \label{MetricH2zzb}
 ds^2 = \frac{r^2 d{\rm z} d\bar{\rm z}}{\rm (1-\frac{z\bar{z}}{4})^2}
\end{align}
In an alternative coordinate system ${\rm z}=2 \tanh\frac{\sigma}{2}\mathrm{e}^{i \phi} \quad \bar{\rm z}=2 \tanh\frac{\sigma}{2}\mathrm{e}^{-i \phi}$, the metric takes the form
\begin{align}
 ds^2 = r^2(d\sigma^2 + \sinh^2\sigma d\phi^2)
\end{align}

For $T\bar{T}$ CFT in $\mathbb{S}^2$ and $\mathbb{H}^2$, we can use trace relation to show the energy-momentum tensor is traceless in connected two point correlators in the large $c$ limit, so the analysis in Appendix \ref{appa} can be carried over to show two point correlators are determined up to a factor as a function of $\mu$
\begin{align} \label{2ptCorrelatorTTS2}
 \<T(\vec{\rm w})T(\vec{\rm v})\>^c = \frac{c}{2}f(\mu) \frac{1}{\rm (w-v)^4}
\end{align}
for $\mathbb{S}^2$ and
\begin{align} \label{2ptCorrelatorTTH2}
 \<T(\vec{\rm w})T(\vec{\rm v})\>^c = \frac{c}{2}g(\mu) \frac{1}{\rm (w-v)^4}
\end{align}
for $\mathbb{H}^2$. For $\mathbb{S}^2$, the factor can be determined by using the one point correlator of energy-momentum tensor  in the replica sphere obtained in \cite{Donnelly:2018bef} to compute Renyi entropy of antipodal points
\begin{align}
 \<T_{\phi\phi}\>_{(n)} = \frac{2r^2 \sin^2\theta}{\mu}(1-\frac{1+\frac{\mu c}{24\pi r^2}}{\sqrt{1+\frac{\mu c}{24\pi r^2}+\frac{\mu c}{24\pi r^2}(\frac{1}{n^2}-1)\frac{1}{\sin^2\theta}}})
\end{align}
Taking a variation in $n$, the replica number, which can be interpreted as a variation of the metric, we have
\begin{align}
 \frac{\partial}{\partial n} \<T_{\phi\phi}(\vec{x})\>_{(n)} = -\int \sqrt{g_{(n)}(\vec{y})} d^2\vec{y} \<T_{\phi\phi}(\vec{x})T_{\phi\phi}(\vec{y})\>_{(n)}^c g^{\phi\phi}_{(n)}(\vec{\rm y})
\end{align}
Setting $n=1$ we return to the regular sphere, and by plugging in $T_{\phi\phi} = -{\rm z}^2 T_{\rm zz}-{\rm \bar{z}}^2 T_{\rm \bar{z}\bar{z}} + 2{\rm z\bar{z}} T_{\rm z\bar{z}}$ we get 
\begin{align}
 -\frac{c}{12\pi\sqrt{1+\frac{\mu c}{24\pi r^2}}}=-&\int \frac{r^2}{(1+\frac{\rm y\bar{y}}{4})^2} \frac{i}{2}d{\rm y}\wedge d\bar{\rm y} \frac{(1+\frac{\rm y\bar{y}}{4})^2}{r^2 \rm y\bar{y}}\nonumber\\
 &\< (-{\rm x}^2 T_{\rm zz}(\vec{\rm x})-{\bar{\rm x}}^2 T_{\rm \bar{z}\bar{z}}(\vec{\rm x})+2{\rm x\bar{x}} T_{\rm z\bar{z}}(\vec{\rm x}))(-{\rm y}^2 T_{\rm zz}(\vec{\rm y})-{\bar{\rm y}}^2 T_{\rm \bar{z}\bar{z}}(\vec{\rm y})+2{\rm y\bar{y}} T_{\rm z\bar{z}}(\vec{\rm y}))\>^c 
\end{align}
With the known correlator $\<T(\vec{\rm w})T(\vec{\rm v})\>^c = \frac{c}{2}f(\mu) \frac{1}{\rm (w-v)^4}$, and by repeated use of Ward identity of conservation of energy momentum tensor we obtain
\begin{align} \label{2ptTTS2ContactTerm}
\<\Theta(\vec{\rm w})T(\vec{\rm v})\>^c = &\frac{\pi c}{12}f(\mu)\partial_{\rm w}^2 \delta(\vec{\rm w}-\vec{\rm v}) - \frac{\pi c}{6} f(\mu)\partial_{\rm w}(\frac{\rm \bar{w}}{\rm w\bar{w}+4} \delta(\vec{\rm w}-\vec{\rm v}) ) \nonumber\\
\<\bar{T}(\vec{\rm w})T(\vec{\rm v})\>^c = &-\frac{\pi c}{12}f(\mu){\partial_{\rm w} \partial_{\bar{\rm w}} \delta(\vec{\rm w}-\vec{\rm v})} + \frac{\pi c}{6}f(\mu){(\rm \frac{\bar{w}}{w\bar{w}+4}\partial_{\bar{w}}-\frac{w}{w\bar{w}+4}\partial_w)} \delta(\vec{\rm w}-\vec{\rm v}) \nonumber\\
&+ \frac{\pi c}{3}f(\mu){\rm\frac{1}{w\bar{w}+4} }\delta(\vec{\rm w}-\vec{\rm v}) \nonumber\\
\<\Theta(\vec{\rm w})\Theta(\vec{\rm v})\>^c=&-\frac{\pi c}{12} f(\mu)\partial_{\rm w}\partial_{\rm \bar{w}}\delta(\vec{\rm w}-\vec{\rm v}) - \frac{2\pi c}{3}f(\mu)\frac{1}{\rm w\bar{w}+4}\delta(\vec{\rm w}-\vec{\rm v})
\end{align}
where $\delta(\vec{\rm w}-\vec{\rm v})$ is the delta function with respect to the measure $\frac{i}{2}d{\rm v} \wedge d\bar{\rm v}$.
\footnote{Most of the time we only consider correlators at distinct points, but here it's an integrated formula which requires inclusion of contact terms.}
Plugging in these correlators and completing the integration, we finally get
\begin{align}
 f(\mu) = \frac{1}{\sqrt{1+\frac{\mu c}{24\pi r^2}}}
\end{align}
By the same token, we need to work out one point correlator of energy-momentum tensor in the replica hyperbolic space
\begin{align}
 ds^2 = r^2(d\sigma^2 + \sinh^2\sigma n^2 d\phi^2)
\end{align}
to find the factor $g(\mu)$ for $\mathbb{H}^2$. We play the same trick as in \cite{Donnelly:2018bef}, that is, we solve (\ref{TTLargec}) together with the conservation equation
\begin{align}
 \nabla_i \<T^i_j\> = 0
\end{align}
in the replica hyperbolic space with the conical singularity smoothed.
\footnote{We also have to make the same assumption in \cite{Donnelly:2018bef}, that is, the trace relation holds in the replica hyperbolic space and the $T\bar{T}$ operator can still be defined as point splitting product, as least to the first order in the replica number $n$.}
We find
\begin{align}
 \<T_{\phi\phi}\>_{(n)} = \frac{2r^2 \sinh^2\sigma}{\mu}(1-\frac{1-\frac{\mu c}{24\pi r^2}}{\sqrt{1-\frac{\mu c}{24\pi r^2}+\frac{\mu c}{24\pi r^2}(\frac{1}{n^2}-1)\frac{1}{\sinh^2\sigma}}})
\end{align}
Plugging  $\<T(\vec{\rm w})T(\vec{\rm v})\>^c = \frac{c}{2}g(\mu) \frac{1}{\rm (w-v)^4}$ and
\begin{align}
\<\Theta(\vec{\rm w})T(\vec{\rm v})\>^c = &\frac{\pi c}{12}g(\mu)\partial_{\rm w}^2 \delta(\vec{\rm w}-\vec{\rm v}) - \frac{\pi c}{6} g(\mu)\partial_{\rm w}(\frac{\rm \bar{w}}{\rm w\bar{w}-4} \delta(\vec{\rm w}-\vec{\rm v}) ) \nonumber\\
\<\bar{T}(\vec{\rm w})T(\vec{\rm v})\>^c = &-\frac{\pi c}{12}g(\mu){\partial_{\rm w} \partial_{\bar{\rm w}} \delta(\vec{\rm w}-\vec{\rm v})} + \frac{\pi c}{6}g(\mu){(\rm \frac{\bar{w}}{w\bar{w}-4}\partial_{\bar{w}}-\frac{w}{w\bar{w}-4}\partial_w)} \delta(\vec{\rm w}-\vec{\rm v}) \nonumber\\
&+ \frac{\pi c}{3}g(\mu){\rm\frac{1}{w\bar{w}-4} }\delta(\vec{\rm w}-\vec{\rm v}) \nonumber\\
\<\Theta(\vec{\rm w})\Theta(\vec{\rm v})\>^c=&-\frac{\pi c}{12} g(\mu)\partial_{\rm w}\partial_{\rm \bar{w}}\delta(\vec{\rm w}-\vec{\rm v}) + \frac{2\pi c}{3}g(\mu)\frac{1}{\rm w\bar{w}-4}\delta(\vec{\rm w}-\vec{\rm v})
\end{align}
into
\begin{align}
 \frac{\partial}{\partial n} \<T_{\phi\phi}(\vec{x})\>_{(n)}|_{n=1} = -\int \sqrt{g(\vec{y})} d^2\vec{y} \<T_{\phi\phi}(\vec{x})T_{\phi\phi}(\vec{y})\>^c g^{\phi\phi}(\vec{\rm y})
\end{align}
we get
\begin{align}
 g(\mu)=\frac{1}{\sqrt{1-\frac{\mu c}{24 \pi r^2}}}
\end{align}

Now we compute three point correlators in $\mathbb{S}^2$. Using the trace relation
\begin{align} \label{TraceRelationS2}
 \Theta(\vec{\rm z}) = -\frac{\mu}{4\pi r^2}(1+\frac{\rm z\bar{z}}{4})^2(T(\vec{\rm z})\bar{T} (\vec{\rm z})- \Theta(\vec{\rm z})^2)
\end{align}
we have
\begin{align} \label{3ptCorrelatorTTS2}
 \<T(\vec{\rm \zeta})\Theta(\vec{\rm z})\bar{T}(\vec{\rm w})\>^c &= -\frac{\mu}{4\pi r^2}(1+\frac{\rm z\bar{z}}{4})^2\<T(\vec{\rm \zeta})(T(\vec{\rm z})\bar{T}(\vec{\rm z})-\Theta(\vec{\rm z})^2)\bar{T}(\vec{\rm w})\>^c \nonumber\\
 &= -\frac{\mu}{4\pi r^2}(1+\frac{\rm z\bar{z}}{4})^2(\<T(\vec{\rm \zeta})T(\vec{\rm z})\>^c\<\bar{T}(\vec{\rm z})\bar{T}(\vec{\rm w})\>^c - 2\<T(\vec{\rm \zeta})\Theta(\vec{\rm z})\bar{T}(\vec{\rm w})\>^c\<\Theta(\vec{\rm z})\>)
\end{align}
Plugging in $\<\Theta(\vec{\rm z})\> = \frac{2\pi r^2}{\mu}(1-\sqrt{1+\frac{\mu c}{24\pi r^2}})\frac{1}{(1+\frac{\rm z\bar{z}}{4})^2}$ obtained from (\ref{1ptCorrelatorTTS2}), we find
\begin{align}
 \<T(\vec{\rm \zeta})\Theta(\vec{\rm z})\bar{T}(\vec{\rm w})\>^c &= -\frac{\mu}{4\pi r^2 \sqrt{1+\frac{\mu c}{24 \pi r^2}}}(1+\frac{\rm z\bar{z}}{4})^2 \<T(\vec{\rm \zeta})T(\vec{\rm z})\>^c \<\bar{T}(\vec{\rm z})\bar{T}(\vec{\rm w})\>^c \nonumber\\
 & = -\frac{\mu c^2}{16\pi r^2}(1+\frac{\mu c}{24 \pi r^2})^{-\frac{3}{2}}(1+\frac{\rm z\bar{z}}{4})^2 \frac{1}{\rm (\zeta-z)^4(\bar{z}-\bar{w})^4}
\end{align}
Similarly for $\mathbb{H}^2$ we get
\begin{align} \label{2pt3ptCorrelatorTTH2}
\<T(\vec{\rm \zeta})\Theta(\vec{\rm z})\bar{T}(\vec{\rm w})\>^c &= -\frac{\mu}{4\pi r^2 \sqrt{1-\frac{\mu c}{24 \pi r^2}}}(1-\frac{\rm z\bar{z}}{4})^2 \<T(\vec{\rm \zeta})T(\vec{\rm z})\>^c \<\bar{T}(\vec{\rm z})\bar{T}(\vec{\rm w})\>^c \nonumber\\
 & = -\frac{\mu c^2}{16\pi r^2}(1-\frac{\mu c}{24 \pi r^2})^{-\frac{3}{2}}(1-\frac{\rm z\bar{z}}{4})^2 \frac{1}{\rm (\zeta-z)^4(\bar{z}-\bar{w})^4}
\end{align}

\section{Correlators of energy-momentum tensor of Einstein gravity in cutoff $\rm AdS_3$}
\setcounter{equation}{0}
\label{sec4}
In this section we compute correlators of energy-momentum tensor of Einstein gravity in cutoff $\rm AdS_3$. In the holographic setup, the large $c$ partition function of the $T\bar{T}$ CFT living on the cutoff surface as the boundary of the bulk gravity, as a functional of the boundary metric $h$, is related to the on-shell action of the gravity by
\begin{align} \label{hlgPartitionFunction}
 \log Z[h] = - I_{\rm on-shell}[h]
\end{align}
The action for the Euclidean Einstein gravity is
\begin{align} \label{EinsteinGravityAction}
 I = -\frac{1}{16\pi G}\int_{\cal M} dV (R+\frac{2}{l^2}) - \frac{1}{8\pi G}\int_{\partial{\cal M}} d\sigma K + \frac{1}{8\pi G}\int_{\partial{\cal M}} d\sigma (\frac{1}{l} + \ldots)
\end{align}
The first term is the Einstein-Hilbert action, the second term is the Gibbons-Hawking term where $K=h^{ij}K_{ij}$ is the trace of the extrinsic curvature $K_{ij}$ on the boundary surface, and the third term is the counter term with other possible addition of local functions of the boundary metric omitted. Taking a functional derivative of (\ref{hlgPartitionFunction}) with respect to the boundary metric, we get one point correlator of energy-momentum tensor in $T\bar{T}$ CFT on the left hand side, and the Brown-York tensor on the right hand side
\begin{align} \label{BYTensor}
 \<T_{ij}\>={T^{BY}}_{ij}=\frac{1}{8\pi G}(K_{ij}-K h_{ij}+ \frac{1}{l}h_{ij}) + \ldots
\end{align}
which depends on the extrinsic curvature and the boundary metric. Multi-point connected correlators of energy-momentum tensor can be computed by taking functional derivative of the one point correlator with respect to the metric
\begin{align} \label{MultiPointCorrelator}
 &\<T_{ij}(\vec{z}) T^{kl}(\vec{w})\>^c = -\frac{2}{\sqrt{h(\vec{w})}}  \frac{\delta \<T_{ij}(\vec{z})\>}{\delta  h_{kl}(\vec{w})} \nonumber\\
 &\<T_{ij}(\vec{z})T^{kl}(\vec{w})T^{mn}(\vec{v})\>^c = \frac{(-2)^2}{\sqrt{h(\vec{w})h(\vec{v})}} \frac{\delta^2 \<T_{ij}(\vec{z})\>}{\delta h_{kl}(\vec{w})\delta h_{mn}(\vec{v})} \nonumber\\
 &\<T_{ij}(\vec{\zeta})T^{kl}(\vec{z})T^{mn}(\vec{w})T^{pq}(\vec{v})\>^c = \frac{(-2)^3}{\sqrt{h(\vec{z})h(\vec{w})h(\vec{v})}} \frac{\delta^2 \<T_{ij}(\vec{\zeta})\>}{\delta h_{kl}(\vec{z})\delta h_{mn}(\vec{w})\delta h_{pq}(\vec{v})} \nonumber\\
 &\ldots
\end{align}
Therefore in order to compute gravity correlators of energy-momentum tensor, we have to compute functional derivatives of the extrinsic curvature with respect to the boundary metric. To this end, we solve the variation of the bulk metric in response to variation of the boundary metric, then compute the extrinsic curvature from the bulk metric. 

To begin with, we gauge-fix the metric to be in Gaussian normal coordinates by diffeomorphism, that is, the radial coordinate is the arclength parameter along the geodesic normal to the cutoff surface. For a variation of the boundary metric $\delta h_{ij}=\epsilon f_{ij}$ where $\epsilon$ is the infinitesimal parameter, the bulk metric takes the form
\begin{align} \label{MetricGaussianNormal}
 ds^2 = d\rho^2 + g_{ij}(\vec{x},\rho) dx^i dx^j
\end{align}
where
\begin{align}
 g_{ij}(\vec{x},\rho) = g^{(0)}_{ij}(\vec{x},\rho) + \epsilon g^{(1)}_{ij}(\vec{x},\rho) + \epsilon^2 g^{(2)}_{ij}(\vec{x},\rho) + \epsilon^3 g^{(3)}_{ij}(\vec{x},\rho) + \ldots
\end{align}
Here $\rho$ is the radial coordinate and $x^i$'s are transverse coordinates. In this gauge there are only three independent components of the metric. At the cutoff surface $\rho=\rho_0$, the extrinsic curvature is given by
\begin{align} \label{ExtCurvatureGaussianNormal}
 K_{ij} = \frac{1}{2}\partial_\rho g_{ij}
\end{align}
The Einstein's equation for the $\rm AdS_3$ gravity is
\footnote{Here we use Greek indices to include both the radial direction and the transverse direction.}
\begin{align}
 R_{\mu\nu} + \frac{2}{l^2} g_{\mu\nu}=0
\end{align}
It's shown in the Appendix \ref{appb} that the Einstein's equation for $\rm AdS_3$ can be decomposed into three equations, the Gauss equation
\begin{align} \label{MainTextGauss}
 K^2- K_{ij}K^{ij} = \hat{R} + \frac{2}{l^2}
\end{align}
the Codazzi equation
\begin{align} \label{MainTextCodazzi}
 \hat{\nabla}^i K_{ij} - \hat{\nabla}_j K = 0
\end{align}
and the radial equation
\begin{align} \label{MainTextRadial}
 \partial_\rho K_{ij} - \frac{1}{2}g_{ij} \partial_\rho K = \frac{1}{2}g_{ij}K^2 - K K_{ij} + 2K_{ik}K_j^k
\end{align}
Solving these three equations order by order, we obtain the Brown York tensor order by order to compute the correlators of energy-momentum tensor. In fact, the Einstein's equation for $\rm AdS_3$ can be further simplified to partial differential equations in the transverse two dimensional space, because the form of the radial dependence of the metric can be solved independently from the boundary metric, following the spirit of \cite{Skenderis:1999nb}. Here we show the results of the gravity correlators and compare them to the correlators in $T\bar{T}$ CFT, leaving details of the computation to Appendix \ref{appc}.

\subsection{$\mathbb{E}^2$ as the cutoff surface}
Pure gravity in Euclidean $\rm AdS_3$ with a cutoff $y=y_0$ in the Poincare patch
\begin{align} \label{MainTextPoincarePatch}
 ds^2 = l^2 \frac{dy^2 + d\vec{x}^2}{y^2}
\end{align}
was proposed to be the holographic dual to $T\bar{T}$ CFT in the cutoff Euclidean plane. In the Appendix \ref{appc}, we computed one point correlators
\begin{align} \label{1ptCorrelatorsAdSE2}
 \<T_{ij}\> = 0
\end{align}
two point correlators
\begin{align} \label{2ptCorrelatorAdSE2}
 \<T(\vec{z})T(\vec{w})\> = \frac{3l}{4G} \frac{1}{(z-w)^4}
\end{align}
three point correlators
\begin{align} \label{3ptCorrelatorAdSE2}
 &\<T(\vec{z})\bar{T}(\vec{w})\bar{T}(\vec{v})\>^c = -\frac{3y_0^2 l}{G}(\frac{1}{(z-w)^3(\bar{w}-\bar{v})^5}+(w \leftrightarrow v)) \nonumber\\
 &\<T(\vec{z})T(\vec{w})T(\vec{v})\>^c=\frac{3l}{2 G}\frac{1}{(z-w)^2(z-v)^2(w-v)^2} \nonumber\\
 &\<T(\vec{z})\Theta(\vec{w})\bar{T}(\vec{v})\>^c = -\frac{9y_0^2 l}{4 G}\frac{1}{(z-w)^4(\bar{w}-\bar{v})^4}
\end{align}
and four point correlators
\begin{align} \label{4ptCorrelatorAdSE2}
 &\<\bar{T}(\vec{\zeta})\Theta(\vec{z})\Theta(\vec{w})\bar{T}(\vec{v})\>^c = \frac{27y_0^4 l}{4G}(\frac{1}{(\bar{\zeta}-\bar{z})^4(z-w)^4(\bar{w}-\bar{v})^4}+(z \leftrightarrow w)) \nonumber\\
 &\<T(\vec{\zeta})\Theta(\vec{z})\bar{T}(\vec{w})\bar{T}(\vec{v})\>^c = -\frac{9y_0^2 l}{2G}\frac{1}{(\zeta-z)^4(\bar{z}-\bar{w})^2(\bar{z}-\bar{v})^2(\bar{w}-\bar{v})^2} \nonumber\\
 &-\frac{9y_0^4 l}{G}\frac{1}{(\zeta-z)^5}(\frac{1}{(\bar{\zeta}-\bar{w})^3(\bar{z}-\bar{v})^4}-\frac{1}{(\bar{z}-\bar{w})^4(\bar{z}-\bar{v})^3} + (w \leftrightarrow v))
\end{align}
After a rescaling of the coordinates $z \rightarrow \frac{y_0}{l} z \quad \bar{z} \rightarrow \frac{y_0}{l} \bar{z}$ to bring the metric in the plane $ds^2 = l^2 \frac{dz d\bar{z}}{y_0^2}$ back to form $ds^2 = dz d\bar{z}$, we find the gravity correlators agree with the $T\bar{T}$ CFT correlators given the holographic dictionary
\begin{align} \label{E2TTAdSDictionary}
 & c= \frac{3l}{2G} \nonumber\\
 & \mu = 16\pi Gl
\end{align}

\subsection{$\mathbb{S}^2$ as the cutoff surface}
Pure gravity in Euclidean $\rm AdS_3$ with a cutoff $\rho=\rho_0$ in the patch
\begin{align} \label{MainTextS2patch}
 ds^2 = l^2(d\rho^2 + \sinh^2\rho(d\theta^2 + \sin^2\theta d\phi^2)) =  l^2(d\rho^2 + \sinh^2\rho\frac{d{\rm z}d\bar{\rm z}}{(1+\frac{\rm z\bar{z}}{4})^2})
\end{align}
is proposed to be the holographic dual to $T\bar{T}$ CFT in the cutoff sphere. We computed 
one point correlator, which is just the Brown-York tensor
\begin{align} \label{1ptCorrelatorAdSS2}
 \<T_{ij}\> = \frac{1}{8\pi Gl}(1-\coth\rho_0) g_{ij}
\end{align}
two point correlators
\begin{align} \label{2ptCorrelatorAdSS2}
 &\<T(\vec{\rm \zeta})T(\vec{\rm z})\>^c = \frac{3l}{4 G \coth\rho_0}\frac{1}{\rm (\zeta-z)^4} \nonumber\\
 &\<T(\vec{\rm \zeta})\bar{T}(\vec{\rm z})\>^c = 0\nonumber\\
 &\<T(\vec{\rm\zeta})\Theta(\vec{\rm z})\>^c = 0 \nonumber\\
 &\<\Theta(\vec{\rm\zeta})\Theta(\vec{\rm z})\>^c = 0
\end{align}
and three point correlators
\begin{align} \label{3ptCorrelatorAdSS2}
 &\<T(\vec{\rm \zeta})\Theta(\vec{\rm z})\bar{T}(\vec{\rm w})\>^c = -\frac{9l\sinh\rho_0}{4 G\cosh^3\rho_0}(1+\frac{\rm z\bar{z}}{4})^2 \frac{1}{\rm(\zeta-z)^4(\bar{z}-\bar{w})^4} \nonumber\\
 &\<T(\vec{\rm \zeta})\bar{T}(\vec{\rm z})\bar{T}(\vec{\rm w})\>^c = \frac{3l\sinh\rho_0}{16 G\cosh^3\rho_0}\rm[\frac{1}{(\bar{z}-\bar{w})^5}(-\frac{\bar{z}\bar{w}}{\zeta-z}+\frac{z\bar{z}\bar{w}+2(\bar{z}+\bar{w})}{(\zeta-z)^2}-\frac{(z\bar{z}+4)(z\bar{w}+4)}{(\zeta-z)^3}) \nonumber\\
 &+(\rm z \leftrightarrow w)] \nonumber\\
 &\<\bar{T}(\vec{\rm \zeta})\bar{T}(\vec{\rm z})\bar{T}(\vec{\rm w})\>^c = \frac{3l(3+\tanh^2\rho_0)\tanh\rho_0}{8 G} \frac{1}{\rm (\bar{\zeta}-\bar{z})^2(\bar{z}-\bar{w})^2(\bar{w}-\bar{\zeta})^2}
\end{align}
We find the gravity correlators agree with the $T\bar{T}$ CFT correlators given the dictionary
\footnote{We only compare correlators computed on both sides. In particular we don't know how to compute $\<\bar{T}(\vec{\rm \zeta})\bar{T}(\vec{\rm z})\bar{T}(\vec{\rm w})\>^c$ for $T\bar{T}$ CFT in $\mathbb{S}^2$ and $\mathbb{H}^2$.}
\begin{align} \label{TTS2AdSDictionary}
 & c= \frac{3l}{2G} \nonumber\\
 & \mu = 16\pi Gl
\end{align}
which takes the same form as $T\bar{T}$ CFT in a Euclidean plane. The sphere has its intrinsic scale $r$, so the second line can also be replaced by
\begin{align}
 \frac{\mu c}{24 \pi r^2} = \frac{1}{\sinh^2 \rho_0}
\end{align}
which relates $T\bar{T}$ deformation parameter to the location of the bulk cutoff.

\section{Summary and discussion}
\setcounter{equation}{0}
\label{sec6}
In this article we have computed large $c$ correlators of energy-momentum tensor for $T\bar{T}$ CFT in a Euclidean plane, a sphere and a hyperbolic space using an operator identity version of the trace relation. To examine the cutoff $\rm AdS$ holographic proposal by Mezei et al. \cite{McGough:2016lol}, we have computed correlators in pure Einstein gravity in Euclidean $\rm AdS_3$ cut off by the Euclidean plane and the sphere, and found agreement with the $T\bar{T}$ CFT correlators given the same dictionary for both cases relating gravity parameters $G,l$ to $T\bar{T}$ CFT parameters $c,\mu$. The cutoff $\rm AdS$ picture was derived from first principle by Guica et al. \cite{Guica:2019nzm} as a pure gravity special case of more general holographic description as $\rm AdS_3$ gravity with mixed boundary condition, for $T\bar{T}$ CFT in a Euclidean plane in the large $c$ limit. Our computation suggests a generalization of Guica's derivation to the case of a sphere. For further research it's also natural to consider correlators of other operators dual to matter fields added to the bulk, and examine the more general holographic description.

Apart from holography, $T\bar{T}$ CFT in a sphere and hyperbolic space deserves further study in its own right. $T\bar{T}$ deformation in a Euclidean plane was shown to be an integrable deformation, but the holographic proposal by Mezei et al. \cite{McGough:2016lol}, the work on partition function and entanglement entropy in \cite{Donnelly:2018bef} and our computation of two point correlators of energy-momentum tensor seems to indicate that large $c$ $T\bar{T}$ flows to trivial in a sphere. On the other hand, correlators of energy-momentum tensor in $T\bar{T}$ CFT in the hyperbolic space blow up and run into a square root singularity when $\mu=\frac{24\pi r^2}{c}$, that may be an indication of failure of the notion of a local energy-momentum tensor. In general, we expect $T\bar{T}$ in curved spaces to be qualitatively different from $T\bar{T}$ in a Euclidean plane in many ways, even though for maximally symmetric spaces the definition of $T\bar{T}$ is somewhat similar. Further study on correlators and entanglement entropy will shed more light on this issue.

We have restricted our work to maximally symmetric spaces. The symmetry does not only greatly reduce the complexity of the computation, but also provides an unambiguous definition of the $T\bar{T}$  operator, assuming the existence of a conserved symmetric energy-momentum tensor. Perhaps the most important open question is to generalize $T\bar{T}$ to generic curved spaces, which has been studied in \cite{Tolley:2019nmm}\cite{Mazenc:2019cfg} and some good results have been obtained, including a derivation of Guica's mixed boundary condition and the large $c$ sphere partition function. It would be interesting to see how the new formalism works at the level of correlators, of energy-momentum tensor and other operators, in and beyond large $c$ limit.

\section*{Acknowledgements}

We would like to thank Ofer Aharony, Per Kraus, Yunfeng Jiang, Ken Kikuchi and  Talya Vaknin for discussion. We also want to express our gratitude to health workers on front line fighting covid-19. YZ is supported by NSFC grant 11905033.

\newpage

\appendix

\section{Maximally symmetric bi-tensor and CFT correlators of energy-momentum tensor in $\mathbb{S}^2$ and $\mathbb{H}^2$}
\setcounter{equation}{0}
\label{appa}

In this appendix we briefly discuss maximally symmetric bi-tensor and derive two point correlators of energy-momentum tensor of CFT in $\mathbb{S}^2$ and $\mathbb{H}^2$, loosely following the notation in \cite{Jiang:2019tcq}.  Roughly speaking, the direction along the geodesic connecting the two points is the only special direction in the (co)tangent spaces of the two points. As a result, it was shown in \cite{Allen:1985wd} that the natural basis for maximally symmetric bi-tensors based on two points $\vec{w}$ and $\vec{v}$ are the operators of parallel transport along the geodesic $I_{ij^{\prime}}(\vec{w},\vec{v})$, the metric at each point $g_{ij}(\vec{w}), \quad g_{k^\prime l^\prime}(\vec{v})$ and the unit tangent vectors to the geodesic at each point $n_i = \partial_{x^i} L(\vec{w},\vec{v}), \quad m_{i^{\prime}} = \partial_{x^{i^{\prime}}}L(\vec{w},\vec{v})$, where $L(\vec{w},\vec{v})$ denotes the geodesic length and the differentiations are with respect to the point $\vec{w}$ and $\vec{v}$, respectively.
\footnote{A word on notation, unprimed indices refer to (co)tangent space at $\vec{w}$ and primed indices refer to (co)tangent space at $\vec{v}$.}
As is shown in \cite{Osborn:1999az}, two point correlator of energy-momentum tensor in a $d$-dimensional maximally symmetric space is a linear combination of five independent bi-tensor structures with coefficients being functions of the geodesic length $L$
\begin{align} \label{2ptCorrelatorBiTensorStructure}
 \<T_{ij}(\vec{w})T_{k^{\prime}l^{\prime}}(\vec{v})\>^{c} &= A_1(L)n_i n_j m_{k^{\prime}} m_{l^{\prime}} \nonumber\\
 &+ A_2(L)(I_{ik^\prime} n_j m_{l^\prime} + I_{i l^\prime} n_j m_{k^\prime} + I_{j k^\prime} n_i m_{l^\prime} + I_{j l^\prime} n_i m_{k^\prime}) \nonumber\\
 & + A_3(L)(I_{i k^\prime}I_{j l^\prime} + I_{i l^\prime}I_{j k^\prime}) + A_4(L)(n_i n_j g_{k^\prime l^\prime}+g_{ij} m_{k^\prime} m_{l^\prime}) \nonumber\\
 &+ A_5(L) g_{ij} g_{k^\prime l^\prime}
\end{align}
This bi-tensor structure is further constrained by conservation of energy-momentum tensor, which by identities
\begin{align}
 &\nabla_i n_j = {\cal A}(g_{ij}-n_i n_j) \nonumber\\
 &\nabla_i m_{j^\prime} = {\cal C}(I_{i j^\prime} + n_i m_{j^\prime}) \nonumber\\
 &\nabla_i I_{j k^\prime} = -({\cal A}+{\cal C})(g_{ij}m_{k^\prime}+I_{i k^\prime}n_j)
\end{align}
reduces to three equations
\begin{align} \label{BiTensorConservation}
 & A_1^{'} - 2 A_2^{'} + A_4^{'} +(d-1)({\cal A}A_1 - 2({\cal A}+{\cal C})A_2) + 2({\cal A}-{\cal C})A_2 + 2{\cal C}A_4 = 0\nonumber\\
 & A_2^{'} - A_3^{'} + d{\cal A}A_3 + {\cal C}A_4 = 0 \nonumber\\
 & A_4^{'} + A_5^{'} + (d-1){\cal A}A_4 + 2{\cal C}A_2 - 2({\cal A}+{\cal C})A_3=0
\end{align}
where
\begin{align}
 {\cal A}(L) = \frac{1}{r}\cot\frac{L}{r}, \quad {\cal C}(L) = -\frac{1}{r}\csc\frac{L}{r}
\end{align}
for sphere, and
\begin{align}
 {\cal A}(L) = \frac{1}{r}\coth\frac{L}{r}, \quad {\cal C}(L) = -\frac{1}{r}\csch \frac{L}{r}
\end{align}
for hyperbolic space. In addition, the second, the third, the fourth and the fifth bi-tensor structures are linearly dependent in two dimensional space, so we can set $A_4=0$ in our cases of $\mathbb{S}^2$ and $\mathbb{H}^2$. For undeformed CFT we assume the energy-momentum tensor is traceless within connected correlators, as a result we get two additional constraints for the correlator
\begin{align} \label{BiTensorTraceless}
 &A_1-4A_2 = 0 \nonumber\\
 &A_3+A_5 = 0
\end{align}
Combining (\ref{BiTensorConservation}) and (\ref{BiTensorTraceless}), we get
\begin{align}
 &A_2=\frac{1}{4} A_1, \quad A_3=-A_5 \nonumber\\
 &\frac{1}{2}A_1^{'}+({\cal A}-{\cal C})A_1=0 \nonumber\\
 &A_5^{'}+2({\cal A}+{\cal C})A_5 + \frac{1}{2} {\cal C}A_1=0
\end{align}
The solution for $\mathbb{S}^2$ is
\begin{align}
 &A_1 = \frac{a_1}{\sin^4\frac{L}{2r}} \nonumber\\
 &A_5 = -\frac{a_1}{8\sin^4\frac{L}{2r}}+\frac{b_5}{\cos^4\frac{L}{2r}} \nonumber\\
 &A_2 = \frac{a_1}{4\sin^4\frac{L}{2r}} \nonumber\\
 &A_3 = \frac{a_1}{8\sin^4\frac{L}{2r}}-\frac{b_5}{\cos^4\frac{L}{2r}}
\end{align}
and the solution for $\mathbb{H}^2$ is
\begin{align}
 &A_1 = \frac{a_1}{\sinh^4\frac{L}{2r}} \nonumber\\
 &A_5 = -\frac{a_1}{8\sinh^4\frac{L}{2r}}+\frac{b_5}{\cosh^4\frac{L}{2r}} \nonumber\\
 &A_2 = \frac{a_1}{4\sinh^4\frac{L}{2r}} \nonumber\\
 &A_3 = \frac{a_1}{8\sinh^4\frac{L}{2r}}-\frac{b_5}{\cosh^4\frac{L}{2r}}
\end{align}
where $a_1$ and $b_5$ are two constants. Because the energy-momentum tensor is symmetric and traceless within connected correlators, it's natural to use the complex stereographic projection coordinates for the sphere, in which the metric takes the form
\begin{align}
 ds^2 = \frac{r^2 d{\rm z} d\bar{\rm z}}{(\rm 1+\frac{z\bar{z}}{4})^2}
\end{align}
and complex Poincare disk coordinates for the hyperbolic space, in which the metric takes the form
\begin{align}
 ds^2 = \frac{r^2 d{\rm z} d\bar{\rm z}}{(\rm 1-\frac{z\bar{z}}{4})^2}
\end{align}
Explicit expressions of ingredients of the bi-tensor structure in these coordinate systems are
\begin{align}
 &L(\vec{\rm w},\vec{\rm v}) = r \cos^{-1} {\rm\frac{w\bar{w}v\bar{v} - 4w\bar{w} - 4v\bar{v} + 8w\bar{v} + 8\bar{w}v +16}{(4+w\bar{w})(4+v\bar{v})}} \nonumber\\
 &I_{{\rm z} {\rm z}^\prime}(\vec{\rm w},\vec{\rm v}) = 0 \nonumber\\
 &I_{\rm z \bar{z}^\prime}(\vec{\rm w},\vec{\rm v}) = {\rm\frac{8r^2(4+\bar{w}v)}{(4+w\bar{w})(4+v\bar{v})(4+w\bar{v})}} \nonumber\\
 &I_{\rm \bar{z} z^\prime}(\vec{\rm w},\vec{\rm v}) = {\rm\frac{8r^2(4+w\bar{v})}{(4+w\bar{w})(4+v\bar{v})(4+\bar{w}v)}} \nonumber\\
 &I_{\rm \bar{z} \bar{z}^\prime}(\vec{\rm w},\vec{\rm v}) = 0
\end{align}
for $\mathbb{S}^2$, and
\begin{align}
 &L(\vec{\rm w},\vec{\rm v}) = r \cosh^{-1} {\rm\frac{w\bar{w}v\bar{v} + 4w\bar{w} + 4v\bar{v} - 8w\bar{v} - 8\bar{w}v +16}{(4-w\bar{w})(4-v\bar{v})}} \nonumber\\
 &I_{{\rm z} {\rm z}^\prime}(\vec{\rm w},\vec{\rm v}) = 0 \nonumber\\
 &I_{\rm z \bar{z}^\prime}(\vec{\rm w},\vec{\rm v}) = {\rm\frac{8r^2(4-\bar{w}v)}{(4-w\bar{w})(4-v\bar{v})(4-w\bar{v})}} \nonumber\\
 &I_{\rm \bar{z} z^\prime}(\vec{\rm w},\vec{\rm v}) = {\rm\frac{8r^2(4-w\bar{v})}{(4-w\bar{w})(4-v\bar{v})(4-\bar{w}v)}} \nonumber\\
 &I_{\rm \bar{z} \bar{z}^\prime}(\vec{\rm w},\vec{\rm v}) = 0
\end{align}
for $\mathbb{H}^2$. Plugging these quantities in \ref{2ptCorrelatorBiTensorStructure}, we find the two point correlators of energy-momentum tensor of CFT in $\mathbb{S}^2$ take the form
\begin{align}
 &\<T^{(0)}_{\rm zz}(\vec{\rm w})T^{(0)}_{\rm zz}(\vec{\rm v})\>^{(0)c} = a_1 r^4 \frac{1}{(\rm w-v)^4} \nonumber\\
 &\<T^{(0)}_{\rm zz}(\vec{\rm w})T^{(0)}_{\rm \bar{z}\bar{z}}(\vec{\rm v})\>^{(0)c} = \frac{1}{2}b_5 r^4 \frac{1}{\rm (1+ \frac{w\bar{v}}{4})^4}
\end{align}
To have the correct flat limit, we must have $a_1 = \frac{c}{8\pi^2 r^4}$ and $b_5=0$, that is
\begin{align}
 &\<T^{(0)}(\vec{\rm w})T^{(0)}(\vec{\rm v})\>^{(0)c} = \frac{c}{2} \frac{1}{(\rm w-v)^4} \nonumber\\
 &\<T^{(0)}(\vec{\rm w})\bar{T}^{(0)}(\vec{\rm v})\>^{(0)c} = 0
\end{align}
Similarly for $\mathbb{H}^2$ we find
\begin{align}
 &\<T^{(0)}(\vec{\rm w})T^{(0)}(\vec{\rm v})\>^{(0)c} = \frac{c}{2} \frac{1}{(\rm w-v)^4} \nonumber\\
 &\<T^{(0)}(\vec{\rm w})\bar{T}^{(0)}(\vec{\rm v})\>^{(0)c} = 0
\end{align}


\section{Geometry of hypersurfaces and Einstein's equation in cutoff $\rm AdS_3$}
\setcounter{equation}{0}
\label{appb}
For self-containedness we offer a basic introduction to the geometry of hypersurface to derive the equations used to compute correlators of energy-momentum tensor in Einstein gravity in cutoff $\rm AdS_3$. A hypersurface $\Sigma$ in a (Pseudo)Riemannian manifold $M$ can be defined as the zero set of a smooth function $\Sigma = \{p \in M, f(p)=0\}$. The canonical normal vector is defined by
\begin{align}
 \zeta = (g^{\mu\nu}\partial_\nu f) \partial_\mu
\end{align}
If $\zeta$ is a null vector, then it's also a tangent vector of the hypersurface. If $\zeta$ is either spacelike or timelike, the tangent space can be decomposed as the direct sum of the tangent space of the hypersurface and the one-dimensional space $N$ spanned by $\zeta$, $T_p M = T_p \Sigma \bigoplus N_p$. In this case we can also define the unit normal $n = \frac{\zeta}{\sqrt{|g(\zeta,\zeta)|}}$ which is normalized to $g(n,n)=\epsilon$ with $\epsilon=1$ for spacelike normal and $\epsilon=-1$ for timelike normal.

Now we consider the extrinsic geometry of the hypersurface. The operator of projection to $T_p\Sigma$, denoted simply by $P$, takes the form in the coordinate basis
\begin{align}
 P^\mu_\nu = \delta^\mu_\nu - \epsilon n^\mu n _\nu
\end{align}
The first fundamental form is given by the induced metric
\begin{align}
 \gamma(X,Y) = g(X,Y) = P_{\mu\nu} X^\mu Y^\nu
\end{align}
for $X, Y \in T_p\Sigma$, where $P_{\mu\nu}=g_{\mu\rho}P^\rho_\nu$. The Weingarten map is defined as
\begin{align}
 L: &T_p\Sigma \rightarrow T_p\Sigma \\ \nonumber
 &X \rightarrow \nabla_X n
\end{align}
and the second fundamental form, also known as the extrinsic curvature, is given by
\begin{align}
 K(X,Y) &= \gamma (L(X),Y) = g(\nabla_X n,Y) = -g(n, \nabla_X Y) = -g(n,\nabla_Y X + [X,Y]) \nonumber\\
 &= -g(n,\nabla_Y X) = K(Y,X)
\end{align}
for $X, Y \in T\Sigma$, with the assumption that the connection is Levi-Civita, that is metric compatible
\begin{align}
 \nabla_X g=0
\end{align}
and torsion free
\begin{align}
 T(X,Y) = \nabla_X Y - \nabla_Y X - [X,Y] = 0
\end{align}
An alternative definition of the extrinsic curvature is given by Lie derivative of the metric in the normal direction
\begin{align}
 K(X,Y) = \frac{1}{2} ({\cal L}_n g) (X,Y)
\end{align}
for $X, Y \in T\Sigma$.

To work out the extrinsic curvature in coordinate basis, we have to do projection onto $T \Sigma$ first
\begin{align}
 K(X,Y) = g(L(P X),P Y)
\end{align}
since the coordinate basis doesn't all lie in $T \Sigma$. We find
\begin{align}
 K_{\mu\nu} = \nabla_\mu n_\nu- \epsilon n_\mu n^\rho \nabla_\rho n_\nu
\end{align}

Now we study the relation between the intrinsic and extrinsic geometry of hypersurfaces. A covariant derivative of a vector can be decomposed into a sum of the part in $T_p \Sigma$ and the part in $N_p$
\begin{align}
 \nabla_X Y = P \nabla_X Y + P_N \nabla_X Y = P \nabla_X Y - \epsilon K(X,Y) n
\end{align}
For $X,Y \in T\Sigma$, we define the covariant derivative in the hypersurface as
\begin{align}
 \hat{\nabla}_X Y = P \nabla_X Y
\end{align}
Because the projection operator $P$ commutes with linear combination over $C^\infty(M)$ and tensor product, $\hat{\nabla}$ is also a connection. Furthermore, for $X,Y,Z \in T\Sigma$
\begin{align}
 \hat{\nabla}_X g(Y,Z) = \nabla_X g(Y,Z) = g(\nabla_X Y,Z)+g(X, \nabla_Y Z) = g(\hat{\nabla}_X Y,Z)+g(X, \hat{\nabla}_Y Z)
\end{align}
and 
\begin{align}
 0&=\nabla_X Y - \nabla_Y X - [X,Y]=\hat{\nabla}_X Y - \epsilon K(X,Y) n - (\hat{\nabla}_Y X - \epsilon K(Y,X) n) - [X,Y] \\ \nonumber
 &= \hat{\nabla}_X Y - \hat{\nabla}_Y X - [X,Y]
\end{align}
so $\hat{\nabla}$ is also Levi-Civita. Needless to say, it coincides with the unique Levi-Civita connection we would have derived from the intrinsic geometry, namely the induced metric. It's natural to define the Riemann curvature tensor in the hypersurface
\begin{align}
 \hat{R}(X,Y)Z = \hat{\nabla}_X \hat{\nabla}_Y Z - \hat{\nabla}_Y \hat{\nabla}_X Z - \hat{\nabla}_{[X,Y]} Z
\end{align}
By definition
\begin{align}
 R(X,Y)Z &= \nabla_X \nabla_Y Z - \nabla_Y \nabla_X Z - \nabla_{[X,Y]} Z  \nonumber \\
 &= \nabla_X (\hat{\nabla}_Y Z - \epsilon K(Y,Z)n) - (X \leftrightarrow Y) - \hat{\nabla}_{[X,Y]} Z + \epsilon K([X,Y],Z) n \nonumber \\
 &= \hat{\nabla}_X \hat{\nabla}_Y Z - \epsilon K(X,\hat{\nabla}_Y Z) n - \epsilon X(K(Y,Z)) n - \epsilon K(Y,Z)\nabla_X n \nonumber \\
 &- (X \leftrightarrow Y) - \hat{\nabla}_{[X,Y]} Z + \epsilon K([X,Y],Z) n \nonumber \\
 &= \hat{R}(X,Y)Z - \epsilon K(X,\hat{\nabla}_Y Z) n - \epsilon X(K(Y,Z)) n - \epsilon K(Y,Z)\nabla_X n \nonumber\\
 &- (X \leftrightarrow Y) + \epsilon K([X,Y],Z) n
\end{align}
The decomposition of the equation above into $T_p \Sigma$ and $N_p$ gives us Gauss and Codazzi equation, respectively. For $W \in T_p \Sigma$
\begin{align}
 g(R(X,Y)Z,W) = g(\hat{R}(X,Y)Z,W) - \epsilon K(X,W)K(Y,Z) + \epsilon K(X,Z)K(Y,W)
\end{align}
and
\begin{align}
 g(R(X,Y)Z,n) &= -K(X,\hat{\nabla}_Y Z) - X(K(Y,Z)) + K(Y,\hat{\nabla}_X Z) + Y(K(X,Z))+ K([X,Y],Z) \nonumber \\
 & = -(\hat{\nabla}_X K) (Y,Z) + K(\hat{\nabla}_X Y,Z) + (\hat{\nabla}_Y K) (X,Z) - K(\hat{\nabla}_Y X,Z) \nonumber\\
 &+ K([X,Y],Z) \nonumber \\
 & = -(\hat{\nabla}_X K) (Y,Z) + (\hat{\nabla}_Y K) (X,Z)
\end{align}
Or in coordinate basis
\begin{align} \label{GaussCodazziCdntBasis}
 &\hat{R}_{\rho\sigma\mu\nu} = P^\alpha_\rho P^\beta_\sigma P^\gamma_\mu P^\delta_\nu R_{\alpha\beta\gamma\delta} + \epsilon K_{\mu\rho}K_{\nu\sigma} - \epsilon K_{\mu\sigma}K_{\nu\rho} \nonumber\\
 & \hat{\nabla}_\mu K_{\nu\sigma} - \hat{\nabla}_\nu K_{\mu\sigma} = -P^\alpha_\mu P^\beta_\nu P^\gamma_\sigma R_{\lambda\gamma\alpha\beta} n^\lambda
\end{align}
The Einstein's equation for the $AdS_{d+1}$ gravity takes the form
\begin{align} \label{Einstein}
 R_{\mu\nu} + \frac{d}{l^2} g_{\mu\nu}=0
\end{align}
where $l$ is the $AdS$ radius. 
We choose a Gaussian normal coordinate patch in which the metric takes the form
\begin{align}
 ds^2= d\rho^2 + g_{ij}(\vec{x},\rho) dx^i dx^j
\end{align}
By definition
\begin{align}
 K(X,Y) = \frac{1}{2} ({\cal L}_n g)(X,Y) = n g(X,Y) - g([n,X],Y) - g(X,[n,Y])
\end{align}
Using $n=\partial_\rho$ and setting $X=\partial_i$, $Y=\partial_j$, we find a simple formula for the extrinsic curvature in this coordinate system
\begin{align} \label{ExtCurvatureGaussianNormal}
 K_{ij} = \frac{1}{2} \partial_\rho g_{ij}
\end{align}
By a double contraction the Gauss equation is reduced to
\footnote{In all of our cases the normal of the cutoff surface is spacelike, so $\epsilon$=1.}
\begin{align} \label{gauss}
 K^2- K_{ij}K^{ij} = \hat{R} + \frac{d(d-1)}{l^2}
\end{align}
By a single contraction the Codazzi equation is reduced to
\begin{align} \label{codazzi}
 \hat{\nabla}^i K_{ij} - \hat{\nabla}_j K = 0
\end{align}
To derive the radial equation, we proceed as
\begin{align}
 R_{\rho j \rho i} = &g(\partial_\rho,R(\partial_\rho,\partial_i)\partial_j)= g(\partial_\rho, \nabla_\rho \nabla_i \partial_j - \nabla_i \nabla_\rho \partial_j) \nonumber\\
 &= \partial_\rho (g(\partial_\rho, \nabla_i \partial_j)) - g(\nabla_\rho \partial_\rho, \nabla_i \partial_j) - \partial_i(g(\partial_\rho,\nabla_j \partial_\rho)) + g(\nabla_i \partial_\rho, \nabla_j \partial_\rho) \nonumber\\
 &=-\partial_\rho K_{ij} + g^{kl} K_{ik} K_{jl}
\end{align}
where $R_{\rho j\rho i}$ is computed to be $R_{\rho j\rho i} = R^\rho_{i\rho j} = R_{ij} - R^k_{ikj} = -\frac{d}{l^2} g_{ij} - R^k_{ikj}$. By a single contraction over roman indices of the Gauss equation we have $R^k_{ikj} = \hat{R}_{ij}-K K_{ij} + K_{ik}K_j^k$, so finally we obtain
\begin{align}
 \partial_\rho K_{ij} = \hat{R}_{ij} - K K_{ij} + 2K_{ik}K_j^k + \frac{d}{l^2} g_{ij}
\end{align}
Using the fact that $\hat{R}_{ij}=\frac{\hat{R}}{2}g_{ij}$ in two dimensional space, we eliminate $\hat{R}_{ij}$ to get a radial equation more practical for computation
\begin{align} \label{radial}
 \partial_\rho K_{ij} - \frac{1}{2}g_{ij} \partial_\rho K = \frac{1}{2}g_{ij}K^2 - K K_{ij} + 2K_{ik}K_j^k
\end{align}

We use these three equations (\ref{gauss})(\ref{codazzi})(\ref{radial}), the same set of equations used in \cite{Kraus:2018xrn}, to compute gravity correlators. However further simplifications are possible. Following the spirit of \cite{Skenderis:1999nb}, we can fix the radial dependence of the bulk metric and reduce the Einstein's equation to partial differential equations in the two dimensional transverse space. For three dimensional space, the Einstein's equation (\ref{Einstein}) fixes the metric to be locally $\rm AdS$
\begin{align}
 R_{\rho\sigma\mu\nu} = -(g_{\mu\rho}g_{\nu\sigma}-g_{\mu\sigma}g_{\nu\rho})
\end{align}
We set $l=1$ for simplicity here and from now on in the appendix. Using (\ref{ExtCurvatureGaussianNormal}), the radial equation now reads
\begin{align} \label{radialCfmFlat}
 -g_{ij}+\frac{1}{2}g_{ij}^{''}-\frac{1}{4}g_{ik}^{'} g^{kl} g_{jl}^{'} = 0
\end{align}
where ``$'$" denotes derivative with respect to $\rho$. It's straightforward to verify, by changing to Fefferman-Graham coordinates $\tilde{\rho}=e^{-2\rho}$ that the radial equation and the uncontracted Gauss and Codazzi equation are equivalent to Equation (7),(8) and (9) in \cite{Skenderis:1999nb}. The radial equation can be integrated to give
\begin{align}
  g = \frac{1}{\tilde{\rho}} g_{(0)} + g_{(2)} + \frac{1}{4}\tilde{\rho} g_{(2)} g_{(0)}^{-1} g_{(2)}
\end{align}
so these three equations are further reduced to Equation (15) in \cite{Skenderis:1999nb} as partial differential equations in the two dimensional transverse space. In the standard context of $\rm AdS$/CFT, $g_{(0)}$ as the metric on the conformal boundary is given, we solve for $g_{(2)}$ to compute various holographic physics quantities as we study holographic Weyl anomaly, holographic renormalization etc. \cite{Henningson:1998gx}\cite{deHaro:2000vlm}. In our context of cutoff $\rm AdS$/$T\bar{T}$ CFT, we fix the metric at a finite cutoff surface as a function of $g_{(0)}$ and $g_{(2)}$ , but still three equations for three independent variables.

\section{Perturbative solutions to Einstein gravity in cutoff $\rm AdS_3$ and correlators of energy-momentum tensor}
\setcounter{equation}{0}
\label{appc}

When the cutoff surface is the two dimensional Euclidean plane $\mathbb{E}^2$, it's natural to use Poincare patch for $\rm AdS_3$
\begin{align}
 ds^2 = \frac{dy^2+d\vec{x}^2}{y^2}
\end{align}
with the cutoff surface at $y=y_0$. Consider a variation of the boundary metric
\begin{align}
 h_{ij}(\vec{x}) = \frac{\eta_{ij}}{y_0^2} + \epsilon f_{ij}(\vec{x})
\end{align}
where $\eta$ is the flat metric, which takes the form $\eta_{ij}=\delta_{ij}$ in the Cartesian coordinates and $\eta_{z\bar{z}}=\eta_{\bar{z}z}=\frac{1}{2}\quad \eta_{zz}=\eta_{\bar{z}\bar{z}}=0$ in the complex coordinates. In response to the variation of boundary metric, the bulk metric now takes the form
\begin{align}
 ds^2 = \frac{dy^2}{y^2} + g_{ij}(y,\vec{x}) dx^i dx^j
\end{align}
where
\begin{align}
 g_{ij}(y,\vec{x}) = \frac{\eta_{ij}}{y^2} + \epsilon g^{(1)}_{ij}(y,\vec{x}) + \epsilon^2 g^{(2)}_{ij}(y,\vec{x}) + \ldots
\end{align}
subject to the boundary condition
\begin{align}
 g^{(1)}_{ij}(y_0,\vec{x})=f_{ij}(\vec{x}), \quad g^{(2)}_{ij}(y_0,\vec{x})=0 \ldots
\end{align}
Now we work out $g_{ij}(y,\vec{x})$ order by order by solving the Einstein's equation. We will give explicit formula for computation to the second order, while computation to the third order and higher, heavily aided by Mathematica, is too complicated to show explicit and complete expression.
The inverse of the metric is computed to be
\begin{align}
 g^{ij} = y^2 \eta^{ij} - \epsilon y^4 \eta^{ik}\eta^{jl} g^{(1)}_{kl} + \epsilon^2 y^6 \eta^{ik}\eta^{mn}\eta^{lj}g^{(1)}_{km}g^{(1)}_{nl} - \epsilon^2 y^4 \eta^{ik}\eta^{jl} g^{(2)}_{kl} + \ldots
\end{align}
and the extrinsic curvature is computed to be
\begin{align}
 K_{ij}&=\frac{1}{2}(-y\partial_y) g_{ij} = \frac{1}{y^2}\eta_{ij} - \frac{1}{2}\epsilon y\partial_y g^{(1)}_{ij} - \frac{1}{2}\epsilon^2 y\partial_y g^{(2)}_{ij} +\ldots\nonumber\\
 &= g_{ij} - \epsilon(g^{(1)}_{ij}+\frac{1}{2}y\partial_y g^{(1)}_{ij}) - \epsilon^2 (g^{(2)}_{ij}+\frac{1}{2}y\partial_y g^{(2)}_{ij})+\ldots
\end{align}
Furthermore we have
\begin{align}
 K=g^{ij} K_{ij}= &2 - \epsilon y^2 g^{(1)}_{ij} \eta^{ij} - \frac{1}{2}\epsilon y^3\partial_y g^{(1)}_{ij} \eta^{ij} + \epsilon^2 y^4 g^{(1)}_ij \eta^{jk} g^{(1)}_{kl}\eta^{li} + \frac{1}{2}\epsilon^2 y^5 \partial_y g^{(1)}_ij \eta^{jk} g^{(1)}_{kl}\eta^{li} \nonumber\\
 &- \epsilon^2 y^2 g^{(2)}_{ij} \eta^{ij} - \frac{1}{2}\epsilon^2 y^3\partial_y g^{(2)}_{ij} \eta^{ij} +\ldots
\end{align}
Plugging these quantities into the radial equation (\ref{radial}), to order $O(\epsilon)$ we have the equation for $g^{(1)}$
\begin{align} \label{radial g1}
 (y\partial_y+\frac{1}{2}y\partial_y y\partial_y)(g^{(1)}_{ij}-\frac{1}{2}\eta_{ij}\eta^{kl}g^{(1)}_{kl})=0
\end{align}
and to order $O(\epsilon^2)$ we have the equation for $g^{(2)}$
\begin{align} \label{radial g2}
 &(y\partial_y+\frac{1}{2}y\partial_y y\partial_y)(g^{(2)}_{ij}-\frac{1}{2}\eta_{ij}\eta^{kl}g^{(2)}_{kl}) \nonumber\\
 &- \frac{1}{4} g^{(1)}_{ij} {\rm tr}(y^3\partial_y y\partial_y g^{(1)}\eta^{-1}) + \frac{1}{4} \eta_{ij} {\rm tr}(y^3\partial_y y\partial_y g^{(1)}\eta^{-1}g^{(1)}\eta^{-1}) - \frac{1}{8}y^2 \eta_{ij}({\rm tr}(y\partial_y g^{(1)}\eta^{-1}))^2 \nonumber\\
 &+ \frac{1}{4}\eta_{ij}{\rm tr}(y^3\partial_y g^{(1)}\eta^{-1}y\partial_y g^{(1)}\eta^{-1}) + \frac{1}{4}y^3\partial_y g^{(1)}_{ij}{\rm tr}(y\partial_y g^{(1)}\eta^{-1}) - \frac{1}{2}y^3\partial_y g^{(1)}_{ik}\eta^{kl}g^{(1)}_{lj} \nonumber\\
 &+ \frac{1}{2}y^3\partial_y g^{(1)}_{ij}{\rm tr}(g^{(1)}\eta^{-1}) - y^3\partial_y(g^{(1)}_{ik}\eta^{kl}g^{(1)}_{lj}) + \frac{3}{2}\eta_{ij}{\rm tr}(y^3\partial_y g^{(1)}\eta^{-1}g^{(1)}\eta^{-1}) - \frac{1}{2}\eta_{ij}{\rm tr}(y^3\partial_y g^{(1)}\eta^{-1}){\rm tr}(g^{(1)}\eta^{-1}) \nonumber\\
 &+y^2 g^{(1)}_{ij}{\rm tr}(g^{(1)}\eta^{-1}) -2y^2g^{(1)}_{ik}\eta^{kl}g^{(1)}_{lj} + y^2\eta_{ij}{\rm tr}(g^{(1)}\eta^{-1}g^{(1)}\eta^{-1}) - \frac{1}{2}y^2\eta_{ij}({\rm tr}(g^{(1)}\eta^{-1}))^2 =0
\end{align}
Similarly the Codazzi equation (\ref{codazzi}) yields an $O(\epsilon)$ equation
\begin{align} \label{codazzi g1}
 -\eta^{ik}\partial_k(g^{(1)}_{ij}+\frac{1}{2}y\partial_y g^{(1)}_{ij}) + \partial_j (g^{(1)}_{ik}\eta^{ik}+\frac{1}{2}y\partial_y g^{(1)}_{ik}\eta^{ik}) = 0
\end{align}
and an $O(\epsilon^2)$ equation
\begin{align} \label{codazzi g2}
 &-\eta^{ik}\partial_k(g^{(2)}_{ij}+\frac{1}{2}y\partial_y g^{(2)}_{ij}) + \partial_j (g^{(2)}_{ik}\eta^{ik}+\frac{1}{2}y\partial_y g^{(2)}_{ik}\eta^{ik}) \nonumber\\
 &+y^4\eta^{im}g^{(1)}_{mn}\eta^{nk}\partial_k(1+\frac{1}{2}y\partial_y)g^{(1)}_{ij} +\frac{1}{2}y^4\eta^{ik}\eta^{lm}[(\partial_k g^{(1)}_{mi}+\partial_i g^{(1)}_{km}-\partial_m g^{(1)}_{ki})(1+\frac{1}{2}y\partial_y)g^{(1)}_{jl}-(i\leftrightarrow j)]\nonumber\\
 &-\partial_j(y^4{\rm tr}(g^{(1)}\eta^{-1}g^{(1)}\eta^{-1})+\frac{1}{2}{\rm tr}(y^5\partial_y g^{(1)}\eta^{-1}g^{(1)}\eta^{-1}))= 0
\end{align}
Finally the Gauss equation (\ref{gauss}) gives us to $O(\epsilon)$
\begin{align} \label{gauss g1}
 (2y^2+y^3\partial_y)g^{(1)}_{ij} \eta^{ij} + y^4(\eta^{im}\eta^{jk}-\eta^{ij}\eta^{km})\partial^2_{ij} g^{(1)}_{km}=0
\end{align}
and to $O(\epsilon^2)$
\begin{align} \label{gauss g2}
 &(2y^2+y^3\partial_y)g^{(2)}_{ij} \eta^{ij} + y^4(\eta^{im}\eta^{jk}-\eta^{ij}\eta^{km})\partial^2_{ij} g^{(2)}_{km} \nonumber\\
 &-\frac{1}{4}({\rm tr}(y^3\partial_y g^{(1)}\eta^{-1}))^2 + \frac{1}{4}y^4{\rm tr}(y\partial_y g^{(1)}\eta^{-1}y\partial_y g^{(1)}\eta^{-1}) - y^6{\rm tr}(g^{(1)}\eta^{-1})(\eta^{im}\eta^{jk}-\eta^{ij}\eta^{km})\partial^2_{ij}g^{(1)}_{km} \nonumber\\
 &+\frac{1}{4}y^6(-4\eta^{ij}\eta^{lm}\eta^{kn}+4\eta^{ij}\eta^{kl}\eta^{mn}+3\eta^{il}\eta^{jm}\eta^{kn}-2\eta^{im}\eta^{kl}\eta^{jn}-\eta^{il}\eta^{jk}\eta^{mn})\partial_i g^{(1)}_{jk} \partial_l g^{(1)}_{mn} \nonumber\\
 &- {\rm tr}(y^5\partial_y g^{(1)}\eta^{-1}){\rm tr}(g^{(1)}\eta^{-1}) - y^4 {\rm tr}(g^{(1)}\eta^{-1}g^{(1)}\eta^{-1}) - y^4({\rm tr}(g^{(1)}\eta^{-1}))^2 = 0
\end{align}

Now we solve for $g^{(1)}$ to compute two point correlators of energy-momentum tensor. From (\ref{radial g1}) we see the traceless part of $g^{(1)}_{ij}$ is a linear combination of a constant and a polynomial of degree minus two in $y$, so $g^{(1)}$ takes the form
\begin{align} \label{g1radial}
 g^{(1)}_{ij}(y,\vec{x}) = A_{ij}(\vec{x})+\frac{B_{ij}(\vec{x})}{y^2} + \frac{1}{2} C(y,\vec{x}) \eta_{ij}
\end{align}
where $B_{ij}$ is subject to the constraint $\eta^{ij}B_{ij}=0$, or $B_{z\bar{z}}=B_{\bar{z}z}=0$ in complex coordinates, as well as $A_{ij}$. Plugging this expression into the Codazzi equation (\ref{codazzi g1}) we get
\begin{align} 
 \partial_j (1+\frac{1}{2}y\partial_y)C(y,\vec{x}) = 2\eta^{ik}\partial_k A_{ij}(\vec{x})
\end{align}
Therefore we have
\begin{align}
 (1+\frac{1}{2}y\partial_y)C(y,\vec{x}) = 2\int \eta^{ik} \partial_k A_{ij}(\vec{x}) dx^j =  4\int \partial_{\bar{z}} A_{zz}(z,\bar{z}) dz + \partial_z A_{\bar{z}\bar{z}}(z,\bar{z}) d\bar{z}
\end{align}
with the integrability condition
\begin{align} \label{Aintegrability}
 \epsilon^{mj}\partial_m \eta^{ik} \partial_k A_{ij} = 4i(\partial_{\bar{z}}^2 A_{zz}-\partial_z^2 A_{\bar{z}\bar{z}}) = 0
\end{align}
So the trace part takes the form
\begin{align} \label{g1codazzi}
 C(y,\vec{x}) = \frac{D(\vec{x})}{y^2} + 4\int \partial_{\bar{z}} A_{zz}(z,\bar{z}) dz + \partial_z A_{\bar{z}\bar{z}}(z,\bar{z}) d\bar{z}
\end{align}
where $\int \partial_{\bar{z}} A_{zz}(z,\bar{z}) dz + \partial_z A_{\bar{z}\bar{z}}(z,\bar{z}) d\bar{z}$ represents the primitive  function whose partial derivatives with respect to $z$ and $\bar{z}$ are $\partial_{\bar{z}} A_{zz}(z,\bar{z})$ and $\partial_z A_{\bar{z}\bar{z}}(z,\bar{z})$, respectively. Next by plugging (\ref{g1radial}) into the Gauss equation (\ref{gauss g1}), we get
\begin{align}
 (2+y\partial_y)C(y,\vec{x}) = y^2(\eta^{ij}\eta^{kl}-\eta^{il}\eta^{jk}) \partial^2_{ij} A_{kl}(\vec{x}) + (\eta^{ij}\eta^{kl}-\eta^{il}\eta^{jk}) \partial^2_{ij} B_{kl}(\vec{x}) + \frac{1}{2} y^2 \Box_{\vec{x}} C(y,\vec{x})
\end{align}
where $\Box_{\vec{x}}=\eta^{ij}\partial^2_{ij}=4\partial_z\partial_{\bar{z}}$ is the Laplacian in the two dimensional Euclidean space. With (\ref{g1codazzi}) plugged in, the equation reduces to
\begin{align} \label{g1gauss}
 8\int \partial_{\bar{z}} A_{zz} dz + \partial_z A_{\bar{z}\bar{z}} d\bar{z} + 4(\partial_{\bar{z}}^2 B_{zz}+\partial_z^2 B_{\bar{z}\bar{z}}) - \frac{1}{2}\Box_{\vec{x}}D = 0
\end{align}
or
\begin{align} \label{mainE2g1}
 \partial_z \partial_{\bar{z}} D - 2(\partial_{\bar{z}}^2 B_{zz}+\partial_z^2 B_{\bar{z}\bar{z}}) = 4\int \partial_{\bar{z}} A_{zz} dz + \partial_z A_{\bar{z}\bar{z}} d\bar{z}
\end{align}
The connected two point correlator of energy-momentum tensor is given by
\begin{align} \label{2pt}
 \<T_{ij}(\vec{z}) T^{kl}(\vec{w})>^c = -\frac{2}{\sqrt{h(\vec{w})}}  \frac{\delta \<T_{ij}(\vec{z})\>}{\delta  h_{kl}(\vec{w})} = -\frac{2}{\sqrt{h(\vec{w})}} \frac{1}{8\pi G}(\frac{\delta K_{ij}(\vec{z})}{\delta h_{kl}(\vec{w})} - \frac{\delta K_{mn}(\vec{z})}{\delta h_{kl}(\vec{w})} h^{mn}(\vec{z}) h_{ij}(\vec{z}))
\end{align}
where $h_{ij}(\vec{x})=\frac{\eta_{ij}}{y_0^2}+\epsilon f_{ij}(\vec{x})=\frac{\eta_{ij}}{y_0^2} + \epsilon(A_{ij}(\vec{x}) + \frac{B_{ij}(\vec{x})}{y_0^2} + \frac{1}{2}(\frac{D(\vec{x})}{y_0^2} + 4\int \partial_{\bar{z}} A_{zz}(z,\bar{z}) dz + \partial_z A_{\bar{z}\bar{z}}(z,\bar{z}) d\bar{z}) \eta_{ij})$ is the boundary metric. We have the boundary condition
\begin{align}
 &f_{zz} = A_{zz} + \frac{B_{zz}}{y_0^2}, \quad f_{\bar{z}\bar{z}} = A_{\bar{z}\bar{z}} + \frac{B_{\bar{z}\bar{z}}}{y_0^2} \nonumber\\
 &f_{z\bar{z}} = \frac{D}{4y_0^2} + \int \partial_{\bar{z}} A_{zz} dz + \partial_z A_{\bar{z}\bar{z}} d\bar{z}
\end{align}
Eliminating $B_{ij}$ and $D$ in favor of $f_{ij}$ and $A_{ij}$, the main equation (\ref{mainE2g1}) takes the form
\begin{align}
y_0^2(\partial_{\bar{z}}^2 f_{zz}+\partial_z^2 f_{\bar{z}\bar{z}}-2\partial_z\partial_{\bar{z}}f_{z\bar{z}})=-2\int(\partial_{\bar{z}} A_{zz}dz + \partial_z A_{\bar{z}\bar{z}}d\bar{z}) 
\end{align}
that is
\begin{align} \label{E2g1propagation}
 \partial_{\bar{z}} A_{zz} &= -\frac{y_0^2}{2}(\partial_z^3 f_{\bar{z}\bar{z}}+\partial_z\partial_{\bar{z}}^2 f_{zz} - 2\partial_z^2\partial_{\bar{z}}f_{z\bar{z}}) \nonumber\\
 \partial_{z} A_{\bar{z}\bar{z}} &= -\frac{y_0^2}{2}(\partial_{\bar{z}}\partial_z^2 f_{\bar{z}\bar{z}}+\partial_{\bar{z}}^3 f_{zz} - 2\partial_z\partial_{\bar{z}}^2 f_{z\bar{z}})
\end{align}
Using the formula $\frac{1}{\pi}\partial_{\bar{z}}\frac{1}{z} = \frac{1}{\pi}\partial_{\bar{z}}\frac{1}{z} = \delta^{(2)}(\vec{z})$, the solution to this propagation equation (\ref{E2g1propagation}) is
\begin{align}
 A_{zz}(\vec{w}) &= -\frac{y_0^2}{2\pi} \int d^2v \frac{1}{w-v} (\partial_v^3 f_{\bar{v}\bar{v}}+\partial_v\partial_{\bar{v}}^2 f_{zz} - 2\partial_v^2\partial_{\bar{v}}f_{z\bar{z}})(\vec{v}) \nonumber\\
 &= \frac{3y_0^2}{\pi} \int d^2v \frac{f_{\bar{z}\bar{z}}(\vec{v})}{(w-v)^4} - \frac{y_0^2}{2} \partial_w\partial_{\bar{w}} f_{zz}(\vec{w}) + y_0^2 \partial_w^2 f_{z\bar{z}}(\vec{w}) \nonumber\\
 A_{\bar{z}\bar{z}}(\vec{w}) &= -\frac{y_0^2}{2\pi} \int d^2v \frac{1}{\bar{w}-\bar{v}} (\partial_v^2 \partial_{\bar{v}} f_{\bar{v}\bar{v}}+\partial_{\bar{v}}^3 f_{zz} - 2\partial_v \partial_{\bar{v}}^2 f_{z\bar{z}})(\vec{v}) \nonumber\\
 &= \frac{3y_0^2}{\pi} \int d^2v \frac{f_{zz}(\vec{v})}{(\bar{w}-\bar{v})^4} - \frac{y_0^2}{2} \partial_w\partial_{\bar{w}} f_{\bar{z}\bar{z}}(\vec{w}) + y_0^2 \partial_{\bar{w}}^2 f_{z\bar{z}}(\vec{w})
\end{align}
where $d^2 v$ is shorthand for $\frac{i}{2}dv\wedge d\bar{v}$. Therefore the variation of the bulk metric in response to the variation of the boundary metric to the first order is
\begin{align} \label{E2g1}
 &g^{(1)}_{ij} = A_{ij} + \frac{B_{ij}}{y^2} + \frac{1}{2}(\frac{D}{y^2}+E)\eta_{ij} \nonumber\\
 &A_{zz}(\vec{w}) = \frac{3y_0^2}{\pi} \int d^2v \frac{f_{\bar{z}\bar{z}}(\vec{v})}{(w-v)^4} - \frac{y_0^2}{2} \partial_w\partial_{\bar{w}} f_{zz}(\vec{w}) + y_0^2 \partial_w^2 f_{z\bar{z}}(\vec{w}) \nonumber\\
 &A_{\bar{z}\bar{z}}(\vec{w}) = \frac{3y_0^2}{\pi} \int d^2v \frac{f_{zz}(\vec{v})}{(\bar{w}-\bar{v})^4} - \frac{y_0^2}{2} \partial_w\partial_{\bar{w}} f_{\bar{z}\bar{z}}(\vec{w}) + y_0^2 \partial_{\bar{w}}^2 f_{z\bar{z}}(\vec{w}) \nonumber\\
 &B_{zz}(\vec{w}) = -\frac{3y_0^4}{\pi}\int d^2v \frac{f_{\bar{z}\bar{z}}(\vec{v})}{(w-v)^4} + y_0^2 f_{zz}(\vec{w}) + \frac{y_0^4}{2}\partial_w\partial_{\bar{w}}f_{zz}(\vec{w}) - y_0^4 \partial_w^2 f_{z\bar{z}}(\vec{w}) \nonumber\\
 &B_{\bar{z}\bar{z}}(\vec{w}) = -\frac{3y_0^4}{\pi}\int d^2v \frac{f_{zz}(\vec{v})}{(\bar{w}-\bar{v})^4} + y_0^2 f_{\bar{z}\bar{z}}(\vec{w}) + \frac{y_0^4}{2}\partial_w\partial_{\bar{w}}f_{\bar{z}\bar{z}}(\vec{w}) - y_0^4 \partial_{\bar{w}}^2 f_{z\bar{z}}(\vec{w}) \nonumber\\
 &D = 4y_0^2 f_{z\bar{z}} + 2y_0^4(\partial_{\bar{z}}^2 f_{zz} + \partial_z^2 f_{\bar{z}\bar{z}} - 2\partial_z\partial_{\bar{z}} f_{z\bar{z}}) \nonumber\\
 &E = -2y_0^2(\partial_{\bar{z}}^2 f_{zz} + \partial_z^2 f_{\bar{z}\bar{z}} - 2\partial_z\partial_{\bar{z}} f_{z\bar{z}})
\end{align}
The first order perturbative computation is enough to compute two point correlators of energy-momentum tensor. The variation of the extrinsic curvature to the first order takes the form
\begin{align}
 \delta K_{ij}=\delta g_{ij} - \epsilon(1+\frac{1}{2}y\partial_y) g^{(1)}_{ij}|_{y=y_0}=\delta h_{ij}-\epsilon A_{ij} -2\epsilon\eta_{ij}\int (\partial_{\bar{z}} A_{zz}dz + \partial_z A_{\bar{z}\bar{z}}d\bar{z}) 
\end{align}
Plugging into (\ref{2pt}) we get
\begin{align}
 \<T_{ij}(\vec{z}) T^{kl}(\vec{w})\>^c = \frac{1}{4\pi G\sqrt{h(\vec{w})}}(\frac{\delta A_{ij}(\vec{z})}{\delta f_{kl}(\vec{w})} - \frac{\delta A_{mn}(\vec{z})}{\delta f_{kl}(\vec{w})} \eta^{mn} \eta_{ij})
\end{align}
Using (\ref{E2g1}) we find
\begin{align}
 &\<T_{zz}(\vec{z})T_{zz}(\vec{w})\>^c = \frac{3}{16\pi^2 G}\frac{1}{(z-w)^4} \nonumber\\
 &\<T_{\bar{z}\bar{z}}(\vec{z})T_{\bar{z}\bar{z}}(\vec{w})\>^c = \frac{3}{16\pi^2 G}\frac{1}{(\bar{z}-\bar{w})^4}
\end{align}
with other two point correlators being zero. Because one point correlators all vanish, connected two point correlators are equal to two point correlators. Comparing with the standard CFT result
\begin{align}
 \<T(\vec{z})T(\vec{w})\> = \frac{c}{2} \frac{1}{(z-w)^4} \nonumber\\
 \<\bar{T}(\vec{z})\bar{T}(\vec{w})\> = \frac{c}{2} \frac{1}{(\bar{z}-\bar{w})^4}
\end{align}
we get the Brown-Henneaux central charge relation
\begin{align}
 c= \frac{3l}{2G}
\end{align}
where we restored the $AdS$ radius $l$ that was previously set to one in our computation.

To compute three point correlators we need to obtain the bulk metric to the second order. Plugging the expression of $g^{(1)}$ into the radial equation (\ref{radial g2}) we obtain
\begin{align}
 (y\partial_y+\frac{1}{2}y\partial_y y\partial_y)(g^{(2)}_{ij}-\frac{1}{2}\eta_{ij}\eta^{kl}g^{(2)}_{kl})=y^2 E A_{ij}
\end{align}
So $g^{(2)}$ takes the form
\begin{align} \label{g2radial}
 g^{(2)}_{ij}(y,\vec{x}) = G_{ij}(\vec{x}) + \frac{H_{ij}(\vec{x})}{y^2} + \frac{1}{4}y^2 E(\vec{x}) A_{ij}(\vec{x}) + \frac{1}{2} F(y,\vec{x})\eta_{ij}
\end{align}
with $\eta^{ij}G_{ij}=\eta^{ij}H_{ij}=0$. Plugging this expression into the Codazzi equation (\ref{codazzi g2}) we get
\begin{align}
 &\partial_z (1+\frac{1}{2}y\partial_y)F = 4 \partial_{\bar{z}} G_{zz} + P_z + y^2 Q_z \nonumber\\
 &\partial_{\bar{z}} (1+\frac{1}{2}y\partial_y)F = 4 \partial_z G_{\bar{z}\bar{z}} + P_{\bar{z}} + y^2 Q_{\bar{z}}
\end{align}
where
\begin{align}
 &P_z = 8\partial_z A_{\bar{z}\bar{z}} B_{zz} + 4A_{\bar{z}\bar{z}}\partial_z B_{zz} - 4A_{zz}\partial_z B_{\bar{z}\bar{z}} - 2D\partial_{\bar{z}}A_{zz} - 2\partial_{\bar{z}}(EB_{zz}) + \frac{1}{2} \partial_z(DE) \nonumber\\
 &P_{\bar{z}} = 8\partial_{\bar{z}}A_{zz}B_{\bar{z}\bar{z}} + 4A_{zz}\partial_{\bar{z}}B_{\bar{z}\bar{z}} - 4A_{\bar{z}\bar{z}}\partial_{\bar{z}}B_{zz} - 2D\partial_z A_{\bar{z}\bar{z}} - 2\partial_z(EB_{\bar{z}\bar{z}}) + \frac{1}{2}\partial_{\bar{z}}(DE) \nonumber\\
 &Q_z = 4\partial_z(A_{zz}A_{\bar{z}\bar{z}})-2E\partial_{\bar{z}}A_{zz}+ \frac{1}{2}\partial_z (E^2) \nonumber\\
 &Q_{\bar{z}} = 4\partial_{\bar{z}}(A_{zz}A_{\bar{z}\bar{z}}) - 2E\partial_z A_{\bar{z}\bar{z}} + \frac{1}{2}\partial_{\bar{z}} (E^2)
\end{align}
Therefore we have
\begin{align}
 (1+\frac{1}{2}y\partial_y)F = \int (4 \partial_{\bar{z}} G_{zz} + P_z)dz + (4 \partial_z G_{\bar{z}\bar{z}} + P_{\bar{z}})d\bar{z} + y^2 \int Q_z dz + Q_{\bar{z}}d\bar{z}
\end{align}
with the integrability condition
\begin{align} \label{PQintegrability}
 &4 \partial_{\bar{z}}^2 G_{zz} + \partial_{\bar{z}}P_z = 4 \partial_z^2 G_{\bar{z}\bar{z}} + \partial_z P_{\bar{z}} \nonumber\\
 &\partial_{\bar{z}} Q_z = \partial_z Q_{\bar{z}}
\end{align}
The first is an equation for $g^{(2)}$ and the second, which only involves $g^{(1)}$, holds for the solution (\ref{E2g1}) of $g^{(1)}$. Now we get the trace part of $g^{(2)}$
\begin{align} \label{g2codazzi}
 F(y,\vec{x}) = \frac{I(\vec{x})}{y^2} + \int (4 \partial_{\bar{z}} G_{zz} + P_z)dz + (4 \partial_z G_{\bar{z}\bar{z}} + P_{\bar{z}})d\bar{z} + \frac{1}{2} y^2 \int Q_z dz + Q_{\bar{z}}d\bar{z}
\end{align}
Plugging (\ref{g2radial}) into the Gauss equation (\ref{gauss g2}), we find
\begin{align}
 &-(2+y\partial_y)F - 4y^2(\partial_{\bar{z}}^2 G_{zz}+\partial_z^2 G_{\bar{z}\bar{z}}) - 4(\partial_{\bar{z}}^2 H_{zz}+\partial_z^2 H_{\bar{z}\bar{z}}) - y^4(\partial_{\bar{z}}^2(EA_{zz})+\partial_z^2(EA_{\bar{z}\bar{z}})) + 2y^2 \partial_z\partial_{\bar{z}} F \nonumber\\
 &+ y^4R + y^2S + W = 0
\end{align}
where
\begin{align}
 R = &4\partial_z A_{\bar{z}\bar{z}} \partial_{\bar{z}} A_{zz} - 4\partial_{\bar{z}}A_{\bar{z}\bar{z}}\partial_z A_{zz} + 4E\partial_z^2 A_{\bar{z}\bar{z}} + 4E\partial_{\bar{z}}^2 A_{zz} + 2\partial_z E \partial_z A_{\bar{z}\bar{z}} + 2\partial_{\bar{z}}E \partial_{\bar{z}}A_{zz} \nonumber\\
 &- \partial_z E \partial_{\bar{z}}E - 2E\partial_z\partial_{\bar{z}}E  \nonumber\\
   = &\frac{1}{4}\partial_{\bar{z}}E \partial_z E - 4\partial_{\bar{z}}A_{\bar{z}\bar{z}}\partial_z A_{zz} \nonumber\\
 S = &4\partial_z A_{\bar{z}\bar{z}} \partial_{\bar{z}}B_{zz} - 4\partial_{\bar{z}}A_{\bar{z}\bar{z}} \partial_z B_{zz} - 4\partial_z A_{zz} \partial_{\bar{z}} B_{\bar{z}\bar{z}} + 4\partial_{\bar{z}} A_{zz} \partial_z B_{\bar{z}\bar{z}} + 8A_{zz}A_{\bar{z}\bar{z}} + 2\partial_z D \partial_z A_{\bar{z}\bar{z}} + 2\partial_{\bar{z}}D \partial_{\bar{z}}A_{zz} \nonumber\\
 &+ 4D\partial_{\bar{z}}^2 A_{zz} + 4D\partial_z^2 A_{\bar{z}\bar{z}} + 2\partial_z E \partial_z B_{\bar{z}\bar{z}} + 2\partial_{\bar{z}}E \partial_{\bar{z}}B_{zz} + 4E\partial_z^2 B_{\bar{z}\bar{z}} + 4E\partial_{\bar{z}}^2 B_{zz} \nonumber\\
 &- \partial_z D \partial_{\bar{z}} E - \partial_{\bar{z}}D \partial_z E - 2E\partial_z\partial_{\bar{z}}D + -2D\partial_z\partial_{\bar{z}}E + \frac{3}{2}E^2 \nonumber\\
 W = &8A_{\bar{z}\bar{z}}B_{zz}+8A_{zz}B_{\bar{z}\bar{z}} + 4\partial_z B_{\bar{z}\bar{z}}\partial_{\bar{z}}B_{zz} - 4\partial_{\bar{z}}B_{\bar{z}\bar{z}}\partial_z B_{zz} + 2\partial_z B_{\bar{z}\bar{z}}\partial_z D + 2\partial_{\bar{z}}B_{zz}\partial_{\bar{z}}D \nonumber\\
 &+ 4D\partial_z^2 B_{\bar{z}\bar{z}} + 4D\partial_{\bar{z}}^2 B_{zz} - \partial_z D\partial_{\bar{z}}D -2D\partial_z\partial_{\bar{z}}D + DE
\end{align}
Substituting (\ref{g2codazzi}) into the equation above, we obtain
\begin{align}
 &-\partial_z^2(EA_{\bar{z}\bar{z}})-\partial_{\bar{z}}^2(EA_{zz}) + \frac{1}{2}(\partial_{\bar{z}}Q_z + \partial_z Q_{\bar{z}}) + R = 0 \nonumber\\
 &-2\int Q_z dz + Q_{\bar{z}} d\bar{z} + \partial_{\bar{z}} P_z + \partial_z P_{\bar{z}} + S = 0 \nonumber\\
 &-2\int (4\partial_{\bar{z}}G_{zz}+P_z)dz + (4\partial_z G_{\bar{z}\bar{z}}+P_{\bar{z}})d\bar{z} - 4(\partial_z^2 H_{\bar{z}\bar{z}}+\partial_{\bar{z}}^2 H_{zz}) + 2\partial_z\partial_{\bar{z}}I + W = 0
\end{align}
It's straightforward to verify the first two equations hold for the solution of $g^{(1)}$, while the last one, together with the boundary condition $g^{(2)}_{ij}|_{y=y_0} = 0$, reduces to two equations for $g^{(2)}$
\begin{align}
 &4\partial_{\bar{z}}G_{zz} + P_z = \partial_z(\frac{y_0^4}{2}R-\frac{y_0^2}{2}(\partial_{\bar{z}}P_z+\partial_z P_{\bar{z}})+\frac{1}{2}W) \nonumber\\
 &4\partial_z G_{\bar{z}\bar{z}} + P_{\bar{z}} = \partial_{\bar{z}}(\frac{y_0^4}{2}R-\frac{y_0^2}{2}(\partial_{\bar{z}}P_z+\partial_z P_{\bar{z}})+\frac{1}{2}W)
\end{align}
Therefore the solution for $g^{(2)}$ can be written as
\begin{align} \label{g2}
 &g^{(2)}_{ij} = G_{ij} + \frac{H_{ij}}{y^2} + \frac{y^2}{4} E A_{ij} + \frac{1}{2} F \eta_{ij} \nonumber\\
 &G_{zz}(\vec{w}) = \frac{1}{4\pi} \int d^2v \frac{1}{w-v} [\partial_z(\frac{y_0^4}{2}R-\frac{y_0^2}{2}(\partial_{\bar{z}}P_z+\partial_z P_{\bar{z}})+\frac{1}{2}W)-P_z](\vec{v}) \nonumber\\
 &G_{\bar{z}\bar{z}}(\vec{w}) = \frac{1}{4\pi} \int d^2v \frac{1}{\bar{w}-\bar{v}} [\partial_{\bar{z}}(\frac{y_0^4}{2}R-\frac{y_0^2}{2}(\partial_{\bar{z}}P_z+\partial_z P_{\bar{z}})+\frac{1}{2}W)-P_{\bar{z}}](\vec{v}) \nonumber\\
 &H_{zz} = -y_0^2 G_{zz} - \frac{1}{4}y_0^4 E A_{zz} \nonumber\\
 &H_{\bar{z}\bar{z}} = -y_0^2 G_{\bar{z}\bar{z}} - \frac{1}{4}y_0^4 E A_{\bar{z}\bar{z}} \nonumber\\
 &F = (1-\frac{y_0^2}{y^2})(\frac{y_0^4}{2}R-\frac{y_0^2}{2}(\partial_{\bar{z}}P_z+\partial_z P_{\bar{z}})+\frac{1}{2}W)+\frac{1}{2}(y^2-\frac{y_0^4}{y^2})(4A_{zz}A_{\bar{z}\bar{z}}+\frac{1}{4}E^2)
\end{align}
We now use the solution of $g^{(2)}$ and the formula
\begin{align} \label{E23ptCorrelatorFormula}
 \<T_{ij}(\vec{z})T^{kl}(\vec{w})T^{mn}(\vec{v})\>^c &= \frac{(-2)^2}{\sqrt{h(\vec{w})h(\vec{v})}} \frac{\delta^2 \<T_{ij}(\vec{z})\>}{\delta h_{kl}(\vec{w})\delta h_{mn}(\vec{v})} \nonumber\\
 &= \frac{(-2)^2}{\sqrt{h(\vec{w})h(\vec{v})}} \frac{1}{8\pi G}(\frac{\delta^2 K_{ij}(\vec{z})}{\delta h_{kl}(\vec{w})\delta h_{mn}(\vec{v})} - h_{ij}(\vec{z})h^{pq}(\vec{z})\frac{\delta^2 K_{pq}(\vec{z})}{\delta h_{kl}(\vec{w})\delta h_{mn}(\vec{v})})
\end{align}
to compute three point correlators, where the variation of the extrinsic curvature to the second order is given by
\begin{align}
 \delta K_{ij} = \delta g_{ij} - \epsilon(1+\frac{1}{2}y\partial_y)g^{(1)}_{ij}-\epsilon^2(1+\frac{1}{2}y\partial_y)g^{(2)}_{ij}|_{y=y_0}
\end{align}
To compute $\<T(z)\bar{T}(w)\bar{T}(v)\>$, we only turn on $f_{zz}$ while keeping other components of the variation of the boundary metric zero for computational simplicity, and we find from (\ref{g2}) 
\begin{align}
 G_{zz}(\vec{z}) = &\frac{12y_0^6}{\pi^2}\int d^2w\int d^2v \frac{1}{(z-w)^3(\bar{w}-\bar{v})^5} f_{zz}(\vec{w}) f_{zz}(\vec{v}) + \ldots
\end{align}
where we only show terms of the form of double integral of two $f$'s which contribute to the three point correlator. Substituting it into the equation (\ref{E23ptCorrelatorFormula}), we get
\begin{align}
 \<T_{zz}(\vec{z})T^{zz}(\vec{w})T^{zz}(\vec{v})\>^c = -\frac{6y_0^{10}}{\pi^3 G}(\frac{1}{(z-w)^3(\bar{w}-\bar{v})^5}+(w \leftrightarrow v))
\end{align}
or
\begin{align}
 \<T_{zz}(\vec{z})T_{\bar{z}\bar{z}}(\vec{w})T_{\bar{z}\bar{z}}(\vec{v})\>^c = -\frac{3y_0^2}{8\pi^3 G}(\frac{1}{(z-w)^3(\bar{w}-\bar{v})^5}+(w \leftrightarrow v))
\end{align}
or in the more usual normalization
\begin{align}
 \<T(\vec{z})\bar{T}(\vec{w})\bar{T}(\vec{v})\>^c = -\frac{3y_0^2}{G}(\frac{1}{(z-w)^3(\bar{w}-\bar{v})^5}+(w \leftrightarrow v))
\end{align}
Completing computation of other three point correlators in a similar way, we list all non-vanishing and independent three point correlators here
\begin{align}
&\<T(\vec{z})\bar{T}(\vec{w})\bar{T}(\vec{v})\>^c = -\frac{3y_0^2}{G}(\frac{1}{(z-w)^3(\bar{w}-\bar{v})^5}+(w \leftrightarrow v)) \nonumber\\
 &\<T(\vec{z})T(\vec{w})T(\vec{v})\>^c=\frac{3}{2 G}\frac{1}{(z-w)^2(z-v)^2(w-v)^2} \nonumber\\
 &\<T(\vec{z})\Theta(\vec{w})\bar{T}(\vec{v})\>^c = -\frac{9y_0^2}{4 G}\frac{1}{(z-w)^4(\bar{w}-\bar{v})^4}
\end{align}

To compute four point correlators we need to work out the variation of the bulk metric to the third order
\begin{align}
 g_{ij}(y,\vec{x}) = \frac{\eta_{ij}}{y^2} + \epsilon g^{(1)}_{ij}(y,\vec{x}) + \epsilon^2 g^{(2)}_{ij}(y,\vec{x}) + \epsilon^3 g^{(3)}_{ij}(y,\vec{x}) + \ldots
\end{align}
Aided by Mathematica, we find the radial equation for $g^{(3)}$
\begin{align} \label{radial g3}
(y\partial_y+\frac{1}{2}y\partial_y y\partial_y)(g^{(3)}_{ij}-\frac{1}{2}\eta_{ij}\eta^{kl}g^{(3)}_{kl}) + y^2 L_{ij}(\vec{x}) = 0
\end{align}
where $L_{ij}$, as a function of $g^{(1)}$ and $g^{(2)}$, is traceless. Explicit expressions of $L_{ij}$ and other quantities in the third order perturbation are too long to be written down here. From the radial equation we have
\begin{align}
 g^{(3)}(y,\vec{x}) = J_{ij}(\vec{x}) + \frac{K_{ij}(\vec{x})}{y^2} - \frac{1}{4}y^2 L_{ij}(\vec{x}) + \frac{1}{2} N(y,\vec{x})\eta_{ij} 
\end{align}
where $\eta^{ij}J_{ij}=0,\quad \eta^{ij}K_{ij}=0$. Furthermore we find the Codazzi equation to take the form
\begin{align} \label{codazzi g3}
 & \partial_z (1+\frac{1}{2}y\partial_y) N = 4 \partial_{\bar{z}} J_{zz} - 2\Pi_z - 2y^2\partial_{\bar{z}}L_{zz} - 2y^2\Theta_z \nonumber\\
 & \partial_{\bar{z}} (1+\frac{1}{2}y\partial_y)N = 4 \partial_z J_{\bar{z}\bar{z}} - 2\Pi_{\bar{z}} - 2y^2\partial_z L_{\bar{z}\bar{z}} - 2y^2\Theta_{\bar{z}}
\end{align}
where $\Pi_z,\Pi_{\bar{z}},\Theta_z,\Theta_{\bar{z}}$ are functions of $g^{(1)}$ and $g^{(2)}$. So the trace part of $g^{(3)}$ takes the form
\begin{align} \label{g3codazzi}
 N(y,\vec{x}) = &\frac{O(\vec{x})}{y^2} + \int (4 \partial_{\bar{z}} J_{zz} - 2\Pi_z)dz + (4 \partial_z J_{\bar{z}\bar{z}} - 2\Pi_{\bar{z}})d\bar{z} \nonumber\\
 &- y^2 \int (\partial_{\bar{z}}L_{zz} + \Theta_z)dz + (\partial_z L_{\bar{z}\bar{z}} + \Theta_{\bar{z}})d\bar{z}
\end{align}
Finally, the Gauss equation for $g^{(3)}$ reads
\begin{align} \label{gauss g3}
 &-2\int (4 \partial_{\bar{z}} J_{zz} - 2\Pi_z)dz + (4 \partial_z J_{\bar{z}\bar{z}} - 2\Pi_{\bar{z}})d\bar{z} + 4y^2 \int (\partial_{\bar{z}}L_{zz} + \Theta_z)dz + (\partial_z L_{\bar{z}\bar{z}} + \Theta_{\bar{z}})d\bar{z} \nonumber\\
 &+ 2\partial_z\partial_{\bar{z}} O - 2y^2(\partial_{\bar{z}}\Pi_z + \partial_z\Pi_{\bar{z}}) - y^4(\partial_{\bar{z}}\Theta_z + \partial_z\Theta_{\bar{z}}) - 4(\partial_{\bar{z}}^2 K_{zz}+\partial_z^2 K_{\bar{z}\bar{z}}) \nonumber\\
 &+ y^6 X + y^4 Y + y^2 Z + \Omega = 0
\end{align}
where $X,Y,Z,\Omega$ are functions of $g^{(1)}$ and $g^{(2)}$. By counting the powers in $y$, this equation reduces to four equations, three being consistency equations satisfied by the solution of $g^{(1)}$ and $g^{(2)}$, and one being the propagation equation
\begin{align}
 -2\int (4 \partial_{\bar{z}} J_{zz} - 2\Pi_z)dz + (4 \partial_z J_{\bar{z}\bar{z}} - 2\Pi_{\bar{z}})d\bar{z} + 2\partial_z\partial_{\bar{z}} O - 4(\partial_{\bar{z}}^2 K_{zz}+\partial_z^2 K_{\bar{z}\bar{z}}) + \Omega = 0
\end{align}
which reduces to two equations
\begin{align} \label{E2g3propagation}
 &4\partial_{\bar{z}}J_{zz} - 2\Pi_z = \partial_z(\frac{y_0^4}{2}Y+y_0^2(\partial_{\bar{z}}\Pi_z + \partial_z\Pi_{\bar{z}})+\frac{1}{2}\Omega) \nonumber\\
 &4\partial_z J_{\bar{z}\bar{z}} - 2\Pi_{\bar{z}} = \partial_{\bar{z}}(\frac{y_0^4}{2}Y+y_0^2(\partial_{\bar{z}}\Pi_z + \partial_z\Pi_{\bar{z}})+\frac{1}{2}\Omega)
\end{align}
by substituting in the relation
\begin{align}
 &O=-y_0^2\int (4 \partial_{\bar{z}} J_{zz} - 2\Pi_z)dz + (4 \partial_z J_{\bar{z}\bar{z}} - 2\Pi_{\bar{z}})d\bar{z} + y_0^4\int (\partial_{\bar{z}}L_{zz} + \Theta_z)dz + (\partial_z L_{\bar{z}\bar{z}} + \Theta_{\bar{z}})d\bar{z} \nonumber\\
 &K_{zz} = -y_0^2 J_{zz} + \frac{y_0^4}{4} L_{zz} \nonumber\\
 &K_{\bar{z}\bar{z}} = -y_0^2 J_{\bar{z}\bar{z}} + \frac{y_0^4}{4} L_{\bar{z}\bar{z}}
\end{align}
imposed by the boundary condition $g^{(3)}(y_0,\vec{x})=0$. The solution to the propagation equation (\ref{E2g3propagation}) takes the form
\begin{align}
 &J_{zz}(\vec{w}) = \frac{1}{4\pi} \int d^2v \frac{1}{w-v} (\partial_z(\frac{y_0^4}{2}Y+y_0^2(\partial_{\bar{z}}\Pi_z + \partial_z\Pi_{\bar{z}})+\frac{1}{2}\Omega) + 2\Pi_z)(\vec{v}) \nonumber\\
 &J_{\bar{z}\bar{z}}(\vec{w}) = \frac{1}{4\pi} \int d^2v \frac{1}{\bar{w}-\bar{v}} (\partial_{\bar{z}}(\frac{y_0^4}{2}Y+y_0^2(\partial_{\bar{z}}\Pi_z + \partial_z\Pi_{\bar{z}})+\frac{1}{2}\Omega) + 2\Pi_{\bar{z}})(\vec{v})
\end{align}
Using the equation
\begin{align}
 &\<T_{ij}(\vec{\zeta})T^{kl}(\vec{z})T^{mn}(\vec{w})T^{pq}(\vec{v})\>^c = \frac{(-2)^3}{\sqrt{h(\vec{z})h(\vec{w})h(\vec{v})}} \frac{\delta^2 \<T_{ij}(\vec{\zeta})\>}{\delta h_{kl}(\vec{z})\delta h_{mn}(\vec{w})\delta h_{pq}(\vec{v})} \nonumber\\
 &= \frac{(-2)^3}{\sqrt{h(\vec{z})h(\vec{w})h(\vec{v})}} \frac{1}{8\pi G}(\frac{\delta^3 K_{ij}(\vec{\zeta})}{\delta h_{kl}(\vec{z})\delta h_{mn}(\vec{w})\delta h_{pq}(\vec{v})} - h_{ij}(\vec{\zeta})h^{rs}(\vec{\zeta})\frac{\delta^3 K_{rs}(\vec{\zeta})}{\delta h_{kl}(\vec{z})\delta h_{mn}(\vec{w})\delta h_{pq}(\vec{v})})
\end{align}
we find
\begin{align}
 &\<\bar{T}(\vec{\zeta})\Theta(\vec{z})\Theta(\vec{w})\bar{T}(\vec{v})\>^c = \frac{27y_0^4}{4G}(\frac{1}{(\bar{\zeta}-\bar{z})^4(z-w)^4(\bar{w}-\bar{v})^4}+(z \leftrightarrow w)) \nonumber\\
 &\<T(\vec{\zeta})\Theta(\vec{z})\bar{T}(\vec{w})\bar{T}(\vec{v})\>^c = -\frac{9y_0^2}{2G}\frac{1}{(\zeta-z)^4(\bar{z}-\bar{w})^2(\bar{z}-\bar{v})^2(\bar{w}-\bar{v})^2} \nonumber\\
 &-\frac{9y_0^4}{G}\frac{1}{(\zeta-z)^5}(\frac{1}{(\bar{\zeta}-\bar{w})^3(\bar{z}-\bar{v})^4}-\frac{1}{(\bar{z}-\bar{w})^4(\bar{z}-\bar{v})^3} + (w \leftrightarrow v))
\end{align}
In principle one can continue in this way to compute higher point correlators. Computation for the case of a spherical cutoff surface is similar.

\newpage

\end{document}